\documentclass{article}
\usepackage{emulateapj}
\usepackage{apjfonts}
\usepackage{psfig}

\def\etal{{{et al.}}}

\def\rss{r$_{ss}$}

\def\thch{$^{13}$CH}
\def\twch{$^{12}$CH}
\def\teff{T$_{\rm eff}$}
\def\logg{log $g$}
\def\chr{$^{12}$CH/$^{13}$CH }

\begin{document}
\title{Th Ages for Metal-Poor Stars}

\author{Jennifer A. Johnson{\altaffilmark{1}}and Michael Bolte}
\affil{UCO/Lick Observatory, University of California, Santa Cruz,
CA~95064; jennifer@ociw.edu, bolte@ucolick.org}
\altaffiltext{1}{Present address:  OCIW, 813 Santa Barbara St., Pasadena, CA 91101}

\begin{abstract}

With a sample of 22 metal-poor stars, we demonstrate that the
heavy element abundance pattern (Z$\geq 56$) is the same as the r-process
contributions to the solar nebula. This bolsters the results of previous
studies that there is a universal
r-process production pattern.  We use the abundance of thorium in five
metal-poor stars, along with an estimate of the initial Th abundance
based on the abundances of stable r-process elements, to measure their ages.
We have four field red giants with errors of 4.2 Gyr in their ages and
one M92 giant with an error of 5.6 Gyr, based on considering the sources
of observational error only.
We obtain an average age of
11.4 Gyr, which depends critically on the assumption of 
an initial production ratio of Th/Eu of
0.496. If the Universe is 15 Gyr old, then the Th/Eu$_{0}$ should
be 0.590, in agreement with some theoretical models of the r-process.

\end{abstract}


\section{Introduction}

One of the fundamental parameters of the Universe is its age. The
expansion age of the Universe  can be calculated directly from
 $\Omega_{M}$, $\Omega_{\Lambda}$, and H$_{0}$.  Since these quantities
are not easily measured, a lower limit to the age of the Universe from
the ages of the oldest local objects has been an important constraint.
As recently as 1996, the most widely-accepted age of the oldest objects
was larger than the expansion age of the universe for the then-popular
cosmology -- a matter-dominated flat universe with $\Lambda = 0$ and
H$_{0}\sim 70$ (Bolte \& Hogan, 1995).  However, results from
observations of high-z supernovae suggest a non-zero value for
$\Lambda$ and larger expansion ages, 14.2 $\pm$ 1.7 Gyrs (Riess \etal{}
1998) and 14.9 $\pm^{1.4} _{1.1}$ Gyr (Perlmutter \etal{} 1999).
Furthermore, the ages of the globular clusters, the most stringent
local limit, have been revised downward with the lengthening of the Pop
II distance scale after Hipparcos satellite parallaxes were measured for
nearby metal-poor stars.  Carretta \etal{} (2000) combined
results from Hipparcos on the distances to subdwarfs, RRLyrae and
Cepheids to re-calibrate the globular cluster distance scale and found
an average age for the globular clusters of 12.9 $\pm$ 2.9 Gyrs.

However, it would be valuable to have additional methods, independent
of stellar models and the Pop II distant scale, to derive the ages for
old stars. It would also be of interest to measure ages for the field
halo stars, in particular stars with [Fe/H]\footnote{We use the
usual notation [A/B]$\equiv \rm{log}_{10}
(N_A/N_B)_*-\rm{log}_{10}(N_A/N_B)_\odot$ and log$\epsilon(\rm A)\equiv \rm{log}_{10}(N_A/N_H)+12.0$.  A/B indicates $N_A/N_B$.}  
$<-2.5$
-- more metal-poor than the most
chemically deficient globular cluster stars.  Butcher (1987) suggested
using the abundance of the only long-lived isotope of Th, Th-232, and
in particular the Th/Nd ratio, as a method for deriving the ages of
field stars.  With a half-life of 14.05 Gyrs, Th decays over a
cosmologically interesting time. However, Nd is also produced in the
s-process, while Th is a pure r-process product, so the Nd production
may not track the production of Th through Galactic history (Butcher
1987; Mathews \& Schramm 1988).  Pagel (1989) suggested using the Th/Eu
ratio, as the abundance of Eu is dominated by contributions from the
r-process.  Francois, Spite, \& Spite (1993) measured the Th/Eu ratio
in stars with [Fe/H] between $-1$ and $-3$.  They found a fairly flat
ratio, with perhaps a rise at the lowest metallicities, which they
thought implied different chemical evolution histories for these two
elements, though a flat ratio would also be expected if there were no
age-metallicity relation.  Unfortunately, their study was hampered by
unknown blending in the Th region, as were the earlier
investigations of Butcher (1987) and Morell, K\"allander, \& Butcher
(1992).

Sneden \etal{} (1996) analyzed the metal-poor, but heavy-element-rich, 
star CS 22982-052.  They could measure 16 stable elements from Ba
(Z=56) to Os (Z=76), some for the first time in a metal-poor star.
They found that abundances for the 16 elements agreed with a scaled
solar system r-process pattern (\rss).  Th, on the other hand,
was lower than predicted by \rss.  If they assumed that the
deviation from the solar r-process pattern was due to the radioactive
decay of Th, rather than to a lower initial Th abundance in
CS22892-052, they could obtain a lower limit to the age of 15.2 $\pm
$3.4 Gyr.  This paper included a comprehensive list of
transitions in the region of the ThII 4019\AA{} line and the
Th abundance was derived via spectrum synthesis. Westin \etal{}
(2000) measured a Th abundance for another metal-poor giant,
HD115444, and determined ages for it and CS 22892-052 using
theoretical predictions for the production of Th and the stable
elements in the r-process. They found an average age of 15.6 $\pm$ 4
Gyr.

\begin{deluxetable}{lllll}
\tablewidth{0pt} 
\tablenum{1}
\tablecaption{Model Atmosphere Parameters}
\tablehead{
\colhead{star} & \colhead {\teff} & \colhead{\logg} & \colhead{[Fe/H]$_{mod}$}
& \colhead {$\xi$}
}
\startdata
HD 29574  &  4350 & 0.30 & -1.70 & 2.30 \nl 
HD 63791 &   4750 & 1.60 & -1.60 & 1.70 \nl
HD 88609 &   4400 & 0.40 & -2.80 & 2.40 \nl 
HD 108577 & 4900 & 1.10 & -2.20 &  2.10 \nl 
HD 115444 &  4500 & 0.70 & -3.00 & 2.25 \nl 
HD 122563 &  4450 & 0.50 & -2.65 & 2.30 \nl 
HD 126587 &  4675 & 1.25 & -2.90 & 1.90 \nl 
HD 128279 &  5100 & 2.70 & -2.20 & 1.40 \nl
HD 165195 &  4375 & 0.30 & -2.20 & 2.50 \nl
HD 186478 & 4525 & 0.85& -2.40& 2.00 \nl 
HD 216143 &  4500 & 0.70 & -2.10 & 2.10 \nl 
HD 218857 & 4850 & 1.80 & -2.00 & 1.50 \nl 
BD -11 145 &   4650 & 0.70 & -2.30 & 2.00 \nl
BD -17 6036 &   4700 & 1.35&  -2.60 & 1.90 \nl 
BD -18 5550 &  4600 & 0.95 & -2.90 & 1.90 \nl 
BD +4 2621  &  4650 & 1.20 & -2.35 & 1.80 \nl 
BD +5 3098  &  4700 & 1.30 & -2.55 & 1.75 \nl 
BD +8 2856  &  4550 & 0.70 & -2.00 & 2.20 \nl
BD +9 3223 &  5250 & 1.65 & -2.10 & 2.00 \nl 
BD +10 2495 &  4900 & 1.90 & -2.00 & 1.60 \nl 
BD +17 3248 &  5200 & 1.80 & -1.95 & 1.90 \nl 
BD +18 2890 & 4900 & 2.00 & -1.60 & 1.50 \nl 
M92 VII-18 & 4250 & 0.20 & -2.18 & 2.30 \nl    
\enddata
\end{deluxetable}

The 4019\AA{} Th line is weak and is blended with
several lines of other elements. The errors on the Th-based age estimates
are so far dominated by measuring and spectrum synthesis uncertainties
for the Th line. It is therefore possible to reduce the error
in the mean age derived via this method by making the measurement
in additional stars. A second significant source of uncertainty in the
Th-based ages is the assumption that the r-process abundance pattern for 
elements
from Ba to Th is ``universal'' and that the abundance of elements such as 
Ba, Eu, Nd and Sm can be used to estimate the initial Th abundance 
in a star. The consistency of heavy-element abundance ratios in studies to
date supports a universal r-process pattern.
In addition to the spectacular example of CS22892-052,  other observations
of heavy elements (Z$ \geq 56$) in metal-poor stars have in general
agreed with the solar-system r-process pattern.  Sneden \&
Parthasarathy (1983) showed that the metal-poor giant HD 122653 has
heavy element abundances best explained by a pure r-process
contribution that matches \rss.  Gilroy \etal{} (1988) studied 22 stars
with [Fe/H] $< -1.5$ and came to similar conclusions for this larger
sample.  Sneden \etal{} (1998) used GHRS spectra to look at three
elements in the A=195 peak, Os, Ir and Pt, in three metal-poor stars.
Combining these results with ground-based data for other elements, they
again confirmed the universality of \rss{} in this mass range.
In addition to allowing a clean estimate of the initial
Th abundance in stars, this result, if substantiated further, is very important for 
understanding the site of the r-process. 

In order to use elements such as Ba and La to estimate the original Th
abundance and to test for a universal r-process pattern, 
we need to make the assumption that the stars we are studying have no
s-process contributions.  This assumption can be tested since the s-process
produces distinct abundance patterns, such as increased Ba and La but
not Eu, that would be noticeable in abundance ratios.   

To further investigate the nature of r-process abundances and
to estimate ages for additional halo stars based on their Th abundances,
we have obtained high-resolution, high-signal-to-noise spectra of 23
metal-poor stars.  In
five of these stars, the abundance of neutron-capture elements is
high enough that Th is detectable.  We estimate ages for these stars
to provide a lower limit to the age of the Universe independent of
the globular clusters.  We use 22 field stars to test 
the validity of the assumption of universal r-process pattern.

\vspace{.5in}
\section{Observations and Data Reductions}

The stars observed are metal-poor ([Fe/H] $<$ $-1.7$) field giants from
the list of Bond (1980) along with one giant in the globular cluster
M92.  The data were obtained with two instruments. ``HIRES'' is the
echelle spectrograph at the Keck I telescope (Vogt et al., 1994).
These data cover 3200-4700 \AA{}, with R$\sim$ 45,000 and a
signal-to-noise greater than 200 at 4000\AA{} for almost all stars.
HIRES data of 12 stars were obtained in May and June 1997.  The second
instrument was the Hamilton spectrograph on the Shane 3-meter telescope
at Lick Observatory (Vogt, 1987). The Hamilton spectra cover the range
3800-7900 \AA, with a S/N of $\sim$ 100 at 6000 \AA.  We obtained
Hamilton spectra for 10 of the HIRES stars, as well as 11 additional
Bond giants.  The Hamilton spectrograph data were taken to obtain
additional lines in the red, particularly Fe and Ti lines,  to help
refine the model atmosphere parameters.   We also wanted to survey
additional Bond giants to add to the sample of stars with many heavy
elements measured, and to find more stars with high [r-process/Fe] as
possible candidates to measure Th.  The spectra were flat-field
corrected, bias-corrected, extracted and wavelength-calibrated using
IRAF (Tody 1993).  The equivalent widths (EWs) of many light and heavy
elements 
were measured using SPECTRE (Sneden, private
communication).  Abundance analysis was done using MOOG (Sneden 1973).  
The
log of observations can be found in Johnson (2001) (Paper II).  

\begin{figure*}[thb]
\centerline{
\psfig{figure=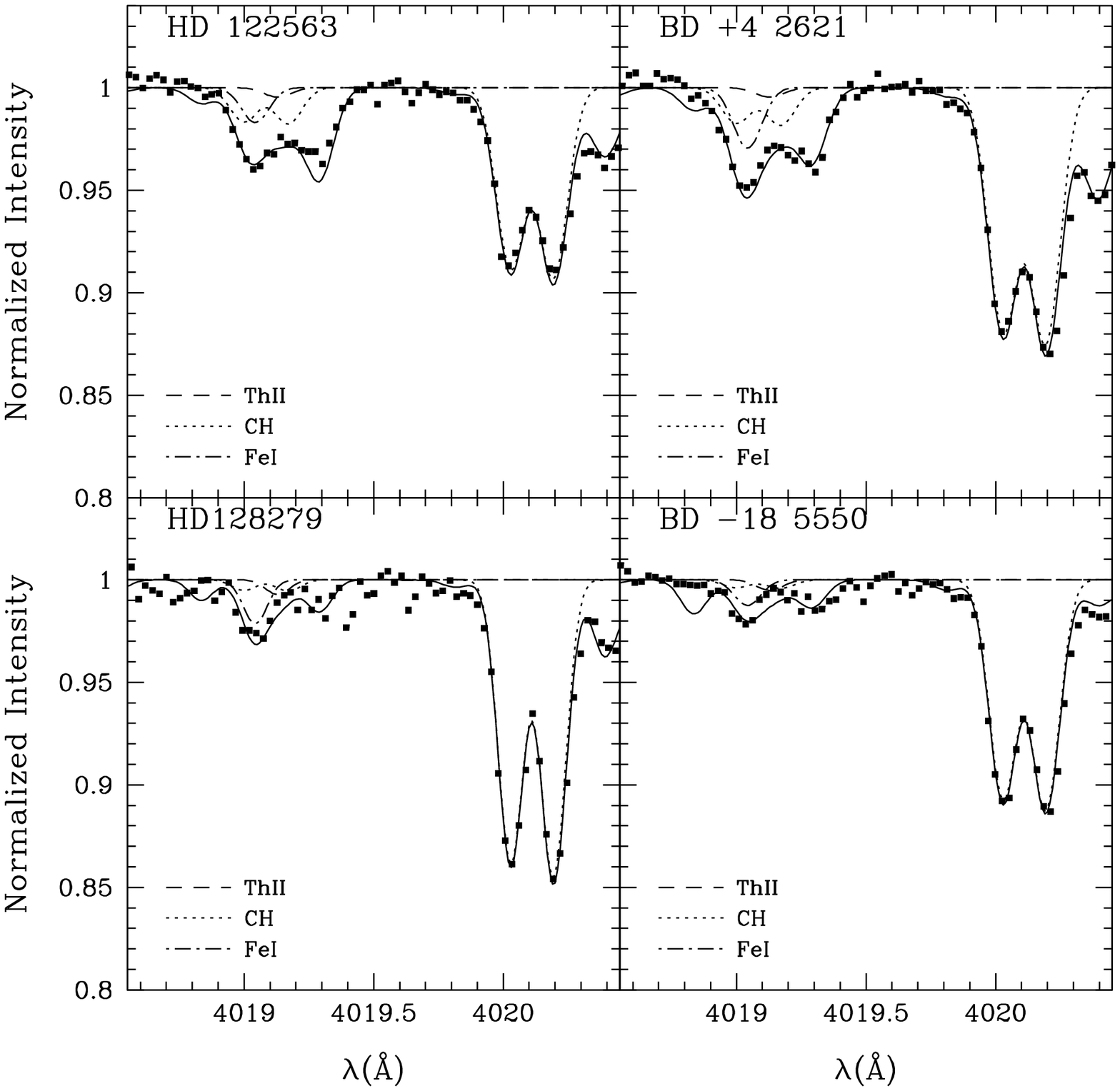,width=7.0truein,angle=0}
}
\caption{Synthesis of the Th region in stars with low [heavy-element/Fe]
ratios.  Only upper limits on the Th abundance could be determined.
The good fit with the observations indicates that our list of lines that
blend with the Th line is reasonably complete and accurate.  Our synthesis
predicts more absorption at 4018.836 \AA{} than is seen because 
our NdII abundances in heavy-element-poor stars are too high.} 
\end{figure*}
\begin{figure*}[htb]
\centerline{
\psfig{figure=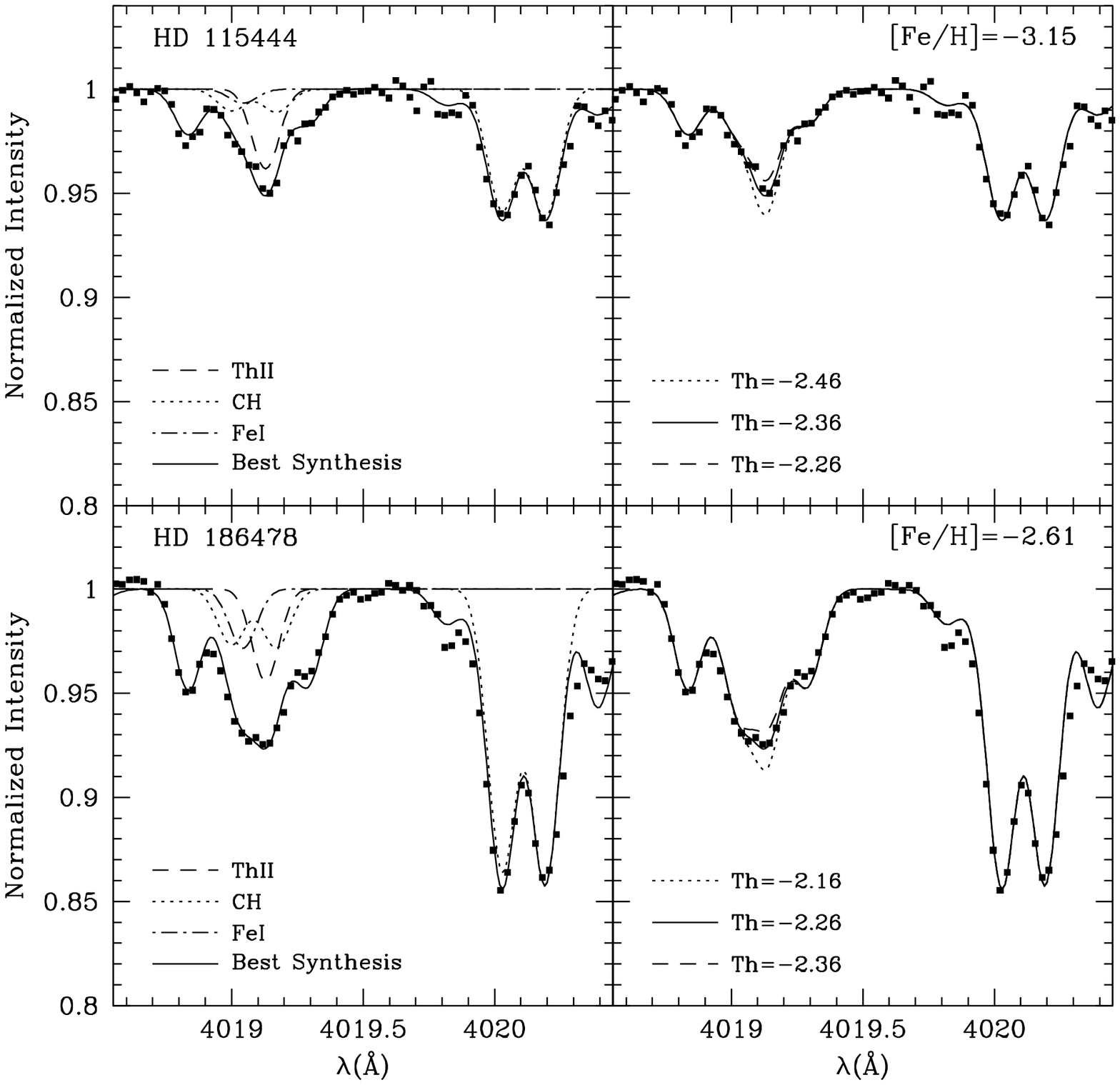,width=7.0truein,angle=0}
}
\caption{Synthesis of the Th region in HD 115444 and HD 186478.  The
left panel shows the overall synthesis with the best-fit Th abundance,
as well as the individual contributions of the major contaminants Fe and \thch.  The right panel shows the best synthesis and the change in the fit if
the Th abundance is changed by $\pm$ 0.10 dex.}
\end{figure*}
\setcounter{figure}{1}
\begin{figure*}[htb]
\centerline{
\psfig{figure=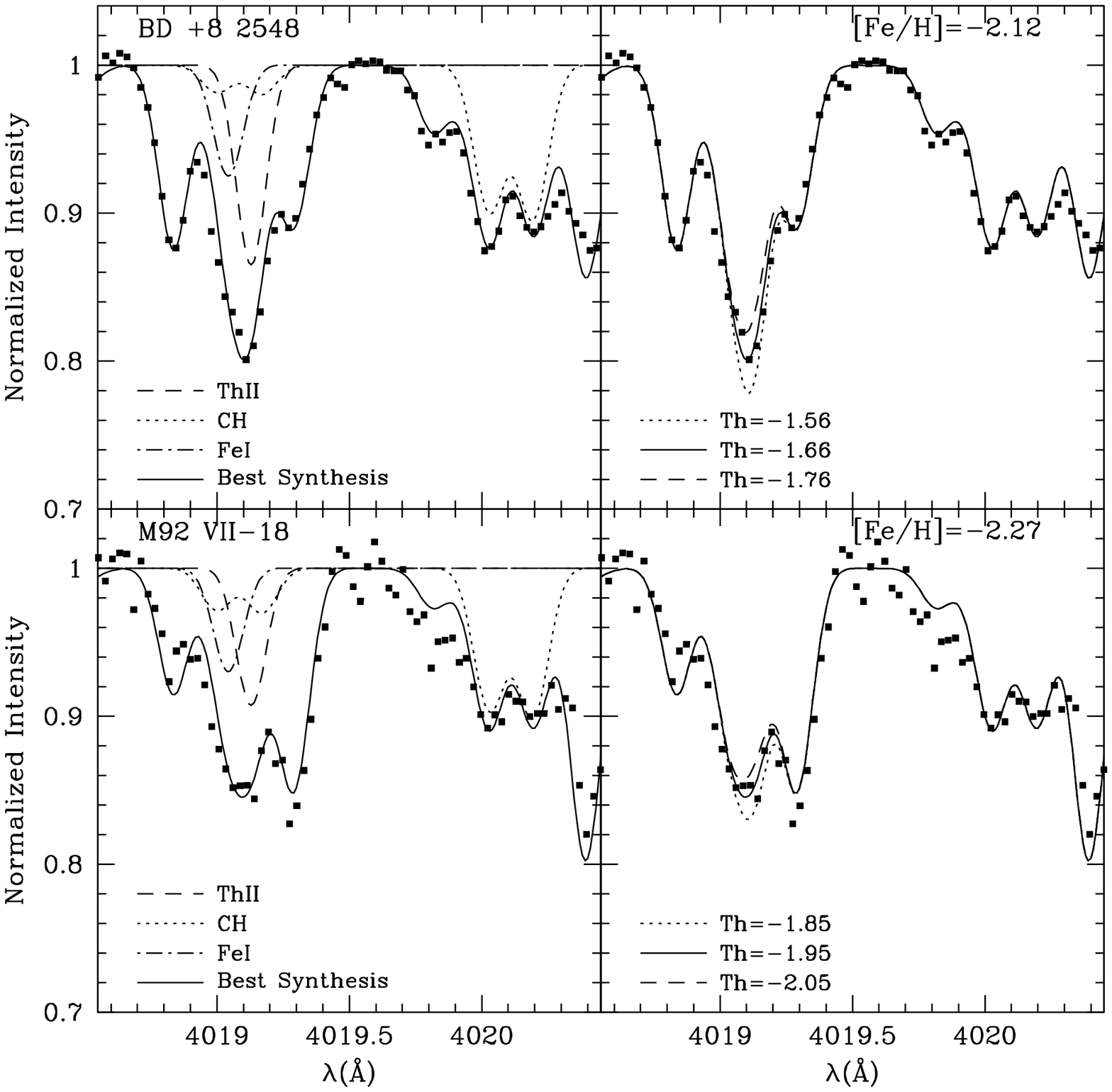,width=7.0truein,angle=0}
}
\caption{Synthesis of the Th region for BD +8 2548 and M92 VII-18.  See
figure caption for Figure 2a.}
\end{figure*}
\setcounter{figure}{1}
\begin{figure*}[htb]
\centerline{
\psfig{figure=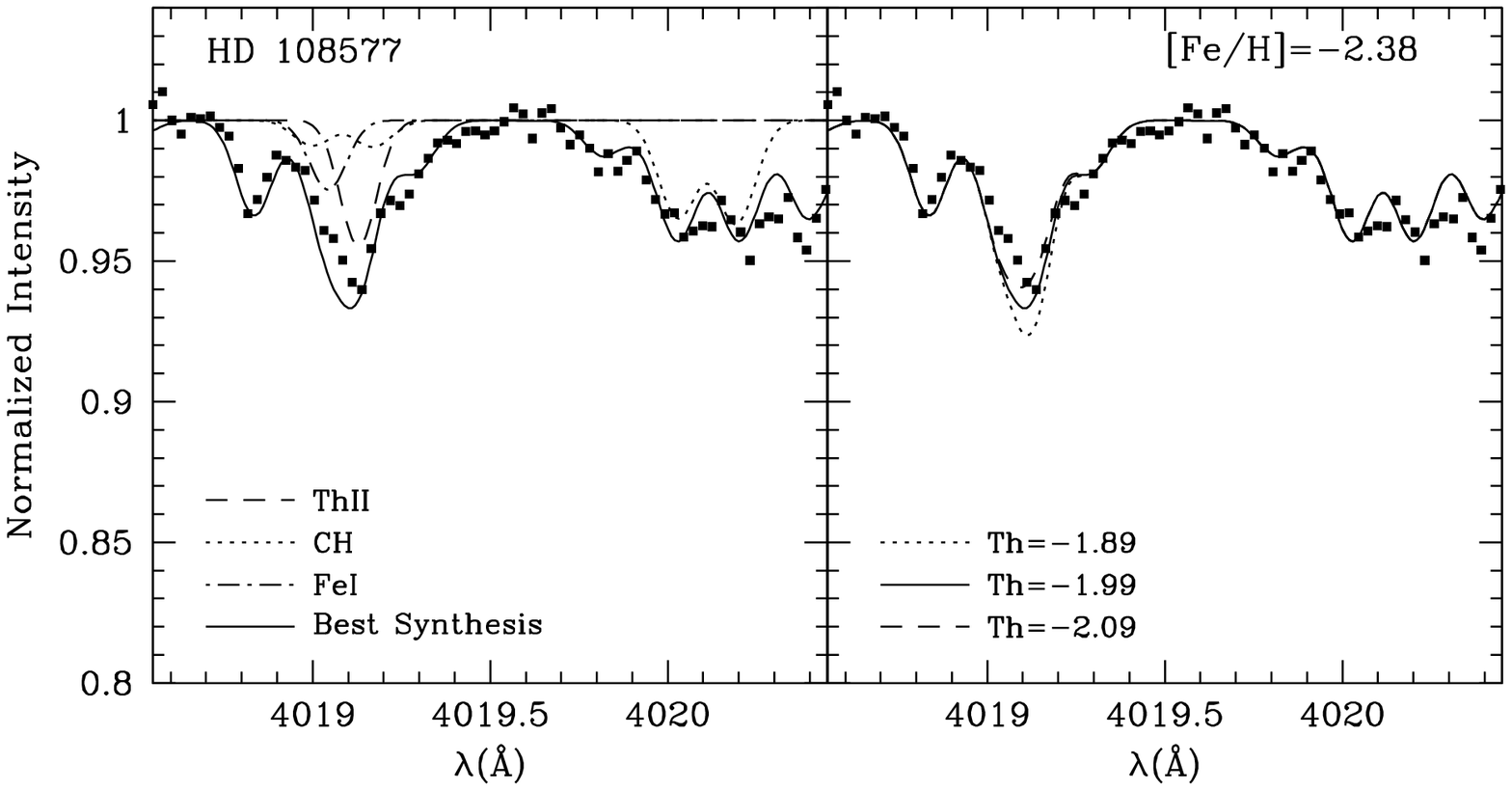,width=7.0truein,angle=0}
}
\caption{Synthesis of the Th region for HD 108577.  See figure caption
for Figure 2a.  The fit to the data would be improved if the FeI abundance
was $\sim$ 0.5 dex lower than the value adopted from EW analysis.  However,
the derived Th abundance would not change.}
\end{figure*}
\setcounter{figure}{2}

\section{Model Atmospheres}

We used the updated model atmospheres of Kurucz (http://cfaku5.harvard.edu/).  Our choices of
model atmosphere parameters are discussed fully in Paper II, and we
summarize the results in Table 1.  In brief, we set the microturbent
velocity ($\xi$) by requiring there be no correlation between the
derived abundance from the CaI, CrI, FeI and TiII lines and their
reduced EWs (RW=EW/$\lambda$). While many of the other elements showed
no trend in abundance as a function of logRW at our adopted $\xi$, some
elements showed trends that changes of up to $\sim \pm 0.3$ km/s in
$\xi$ eliminated.  We have chosen $\pm 0.3$ km/s as our error in
$\xi$.  \teff{} was changed until there was no trend in the abundance
versus excitation potential plot of the FeI lines.  We estimate, based
on the range of \teff{} that produce acceptable fits, that our errors
are $\pm 100$K.  Next, we determined \logg{} by matching the FeI and
FeII abundances. We have only $\sim$ 15 FeII lines and the gf
values for these are generally of lower quality than those for FeI
lines. Our FeII abundances therefore have a relatively large standard
deviation of the mean $\sim$ 0.05.  Also our gravities depend on our
choice of temperature and $\xi$, so our errors in \logg{} are $\pm$ 0.3
dex.  Changing the metallicity of our atmosphere by 0.2 dex only changed the
abundances by $\pm$ 0.01 dex, and therefore that has been ignored
as a source of error.

\section{Abundances}

\subsection{Heavy Elements}

We attempted to determine the abundances for many heavy elements from
Ba (Z=56) to Os (Z=76) in the stars we observed to see if \rss{} was repeated in these stars.  The stars in our sample were more metal-rich than CS22892-052 as
well as less heavy-element rich.  Therefore, blending and detection 
affected some of the lines that Sneden et al. (1996) could use in
their study of CS22892-052.
We included only those lines which were not affected by blending at
the line center, with the
exception of Th (see below).  Excluding blended lines meant we were not able to
measure every element in every star.  In particular, 
for Ho, Hf, and Os, we were able to 
obtain upper limits only, which are still useful in ruling out large ($\sim$
0.5 dex) deviations in the r-process pattern.  Unfortunately, the line
we could use to set limits on the Os (4261.85 \AA) gave a lower
abundance by $\sim$0.50 dex than other lines used by Sneden \etal{} (1996).
for CS 22892-052.
Line parameters and EWs
are listed in Paper II.  Hyperfine splitting was taken into account for
Ba, La, Eu, and Ho.  Based on the abundance pattern seen in the other
elements, we adopted the Ba abundances derived using the solar system
r-process isotopic composition.  Choosing the total solar system
isotopic composition increased the Ba abundances by $\sim 0.02-0.05$.  
Whenever possible, we used analysis of EWs
 of lines to derive abundances of unblended lines.  
Spectral synthesis was used in 
crowded regions.  Our linelists are from Sneden \etal{} (1996).  
Table 2\footnote{Table 2 is included in an appendix} summarizes our heavy element abundances.  For solar values we
have used the photospheric abundances for Anders and Grevesse (1989), 
except for those elements with uncertain photospheric abundances where
meteoritic values were used.  We also adopted log $\epsilon$=7.52 as the solar iron
abundance.

\subsection{Th Abundance}

The only Th line strong enough to measure in our spectra (4019.12
\AA{}) is unfortunately blended with several lines from other
elements.  We made an initial line list based on the atomic data of
Morell \etal{} (1992), Sneden \etal{} (1996) and
Kurucz CD ROM 23.  We then synthesized the solar spectrum using these
lists and adjusted the gf values of lines to match the solar
spectrum.  For crucial lines, we searched the literature for laboratory
values.  For the FeI line, we adopted a log gf of $-2.68$ from May
(1974) and a wavelength of 4019.043 \AA{} from Learner \etal{} (1990).
The Th line also has a wavelength from Learner \etal{} and a laboratory
log gf from Simonsen \etal{} (1990).  The hyperfine $A$ and $B$
constants for the CoI lines at 4019.13 \AA{} and 4019.29 \AA{} are
given in Pickering \& Semeniuk (1995).
Norris, Ryan \& Beers (1997) pointed out the important contribution of
\thch{} lines from the $B \Sigma ^- - X ^2 \Pi$ 0,0 band and suggested
that Kurucz' estimated wavelengths be adjusted by 0.15-0.25 \AA.  We
have a spectrum of the extremely carbon-rich star, CS 22957-057, that
drew Norris \etal's attention to the \thch{} contamination.  This is a
lower S/N ($\sim$ 50) spectrum taken with HIRES during the June run.
Our spectrum confirms the wavelengths for the \thch{} lines found by
Norris \etal{}.  The gf values for the \thch{} lines have been taken
from Kurucz's web site.  Finally, the CeII line can be important in stars with
supersolar [Ce/Fe] ratios.  Sneden \etal{} (1996) found it necessary to
include this line in order to account of the absorption profile in
CS22892-052.  They increased the log gf value by 0.3 dex over the
value from Kurucz' CD-ROM 23.  It made a difference of 0.05 dex in the
derived Th abundance in CS 22892-052.  Here, with better
resolution and smaller CeII abundances, that increase in the Ce II gf
affects the Th abundance at most $\pm$ 0.02.
 Table 3 has the linelist we used, which contains lines up to 1/1000
the strength of the Th line in red giants.

\begin{deluxetable}{cccr} 
\tablenum{3} 
\tablewidth{0pt}
\tablecaption{Linelist near ThII at 4019 \AA} 
\tablehead{ \colhead {$\lambda$ (\AA)} & \colhead {Element} & \colhead {E.P. (eV
)} &
\colhead {log gf} } 
\startdata 
4018.100 & Mn I &   2.11 & -0.309 \nl
4018.266 & Fe I &   3.27 & -1.360 \nl
4018.368 & Zr II &   0.96 & -0.994 \nl 
4018.506 & Fe I &   4.21 & -1.597 \nl 
4018.836 & Nd II & 0.06 & -0.880 \nl 
4018.986 & U II &   0.04 & -1.391 \nl 
4019.000 & $^{13}$CH &   0.46 & -1.163 \nl 
4019.043 & Fe I &   2.61 & -2.680 \nl
4019.057 & Ce II &   1.01 &  0.093 \nl 
4019.067 & Ni I &   1.94 & -3.174 \nl 
4019.110 & Co I &   2.28 & -3.287 \nl 
4019.118 & Co I & 2.28 & -3.173 \nl 
4019.120 & Co I &   2.28 & -3.876 \nl 
4019.125 & Co I &   2.28 & -3.298 \nl 
4019.125 & Co I &   2.28 & -3.492 \nl
4019.130 & Th II &   0.00 & -0.270 \nl 
4019.134 & Co I &   2.28 & -3.287 \nl
4019.135 & Co I &   2.28 & -3.474 \nl 
4019.137 & V I&   1.80 & -1.300 \nl 
4019.138 & Co I &   2.28 & -3.173 \nl 
4019.140 & Co I &   2.28 & -3.298 \nl 
4019.170 & $^{13}$CH &   0.46 & -1.137 \nl 
4019.272 & Co I &   0.58 & -3.480 \nl 
4019.281 & Co I &   0.58 & -3.470 \nl 
4019.294 & Co I &   0.58 & -3.220 \nl 
4019.296 & Co I &   0.58 & -3.330 \nl 
4019.322 & Co I &   0.58 & -4.090 \nl 
4019.332 & Co I &   0.58 & -4.040 \nl 
4019.632 & Pb I &   2.66 & -0.220 \nl 
4019.726 & Gd I &   0.07 & -1.046 \nl 
4019.810 & Nd II &   0.63 & -0.770 \nl 
4019.829 & Sm II & 0.28 & -1.695 \nl 
4019.880 & Fe I &   2.60 & -5.000 \nl 
4019.897 & Ce II &   1.01 & -0.368 \nl 
4019.976 & Sm II &   0.19 & -1.419 \nl 
4020.029 & $^{12}$CH &   0.46 & -1.163 \nl 
4020.051 & Nd II &   1.27 & -0.290 \nl 
4020.193 & $^{12}$CH &   0.46 & -1.137 \nl 
4020.251 & Ni I &   3.70 & -0.936 \nl 
4020.390 & Sc I &   0.00 &  0.039 \nl 
4020.482 & Fe I &   3.64 & -1.900 \nl 
\enddata 
\end{deluxetable}
For every synthesis in the Th region, the Fe, Ni, Nd, and Co
abundances were the abundances previously deduced from the EW
analysis.   The \twch{} lines from the same transition as the 
contaminating \thch{} lines are at 4020 \AA.  
Therefore, the carbon abundance of a star was adjusted
to match the 4020 \AA{} feature.  The \twch/\thch{} ratio was determined using
lines between 4200-4370\AA{} and is listed in Table 4 for stars with Th
abundances.  Because the \thch{} lines in our sample were weak, these
ratios have errors of $\pm 2$.  Ce could only be measured in the 
neutron-capture-rich stars.  For the other stars, our Ce values were
estimated using the Ba abundance and the Ce/Ba ratio found in the 
neutron-capture-rich stars, an assumption which is justified by the results in \S 5.1.

\begin{deluxetable}{cc}
\tablenum{4}
\tablewidth{0pt}
\tablecaption{\chr used for Thorium Synthesis}
\tablehead{\colhead{Star} & \colhead{\chr} }
\startdata
HD 186478  & 6 \nl
HD 115444  & 6 \nl
HD 108577 & 4 \nl
BD -18 5550 & 32 \nl
HD 122563 & 6 \nl
BD +4 2621 & 8 \nl
HD 128279 & 32 \nl
M92 VII-18 & 4 \nl
BD +8 2548 & 6 \nl
\enddata
\end{deluxetable}

We tested the validity of the linelist of contaminants on four stars
with high S/N, but low [heavy-element/Fe] values (Figure 1).  For these
stars we expect little contribution of Th, Nd or Ce to the region.  In
each case, we set the Th to be as large as possible.  These upper
limits are included in Table 2.  The agreement between synthesized and
observed spectra in Figure 1 is encouraging, especially since the amount
of absorption due to Fe and \thch{} varies from star to star.
For the high [heavy-element/Fe] stars,
Figure 2 shows our best synthesis with the individual
contributions of the strongest lines in one panel and the effect of
changing the Th abundance in the second panel.  The Th
 abundance results are included in Table 2.

\subsection{Error Analysis}

A complete discussion of the error analysis can be found in Paper II.  
We consider two sources of error for each element:  line-by-line scatter caused by
errors in gf values and EWs and systematic
offsets caused by incorrect model atmospheres.
For elements with several measurable lines, we estimate the random
errors with the standard deviation of the mean  of
the abundance.  For stars with only one or a few lines of a particular element, 
we established a minimum
standard deviation by looking at the standard deviation for that element in stars with 
many lines or by looking at errors in the EWs or spectral synthesis.  Table 2 lists
the standard deviation of the sample ($\sigma$) for each element, as well as the number of lines used to determine
the abundance.  Also given is the error in [element/H] ($\sigma_{tot}$), 
where the errors associated with the model atmospheres are taken into 
account.

Since only the relative abundances,
rather than absolute [element/H], are important for this method of
determining ages, we performed new 
numerical experiments to calculate the errors in the relative abundances
of the heavy elements as the atmospheric parameters are changed.
For the elements between Z of 56 and 70 as well as Th, we determined the change in the abundance
if we changed the temperature by 100K, log g by 0.3 dex, or $\xi$
by 0.3 km/s. 
The abundances of rare earth elements and Th change in similar directions when
the model atmospheres change, leaving only a small relative difference. 
 The exception is Ba and Yb when $\xi$
is changed, because these elements are usually only represented by much
stronger lines than the rest of the heavy elements.
In addition, in many cases, the abundances change the same
way with an increase in \teff{} and an increase in \logg{}.
Since these quantities are anti-correlated along the red giant branch,
the net effect of changing the model atmosphere parameters is further
decreased.
The
overall error in abundances of the rare earths relative to each other
from the choice of model atmosphere parameters is negligible compared
with the line-by-line scatter.  For example, the error in [Nd/Ce] from
the model atmospheres is $\sim$ 0.01 and for [Sm/Eu] it is $\sim$ 0.05,
smaller than the line-by-line uncertainties of $\sim$ 0.05-0.2 dex.  
With these results in mind, we will use as
errors the standard error of the mean when focusing on these elements.

\begin{figure*}[htb]
\centerline{
\psfig{figure=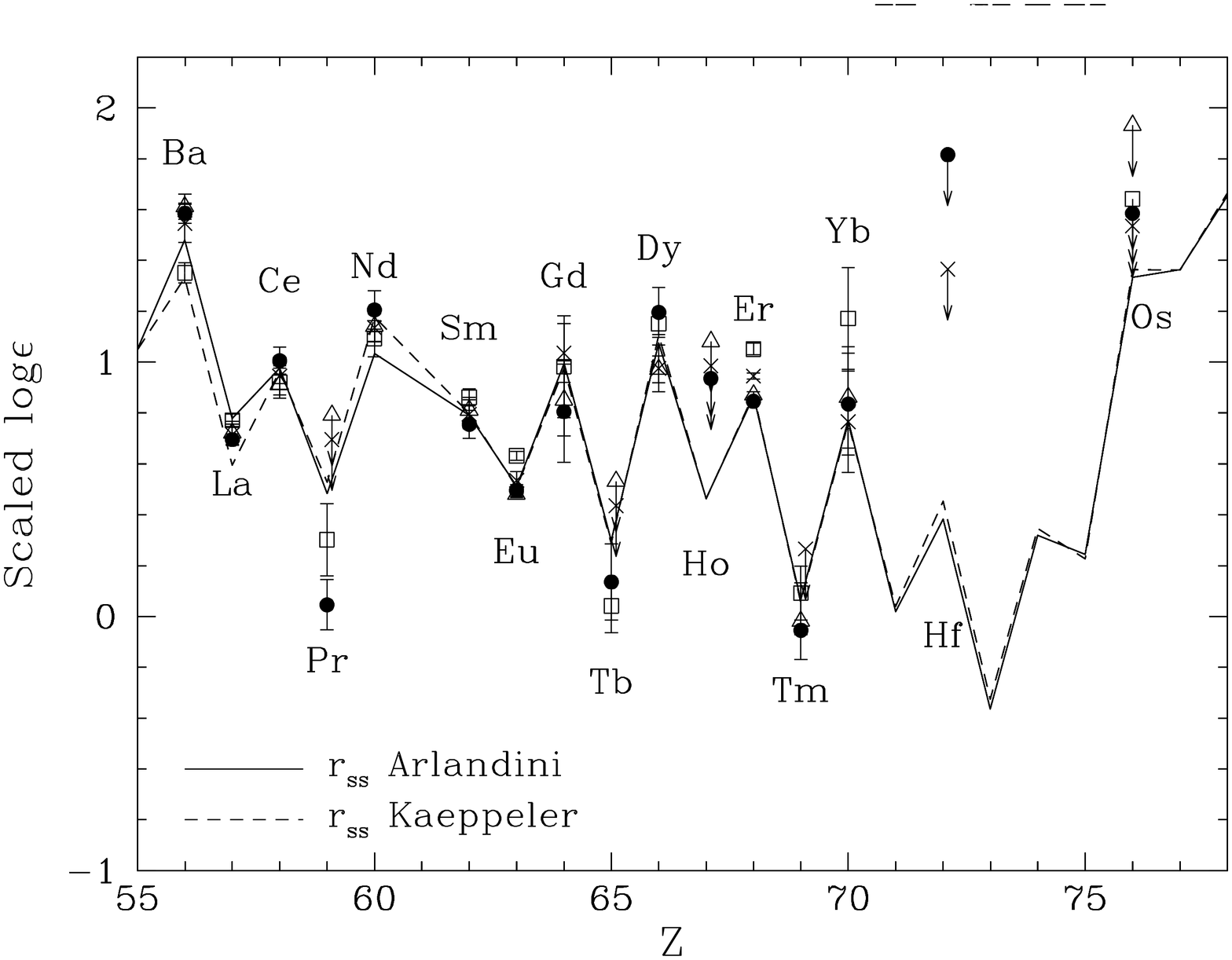,width=7.0truein,angle=270}
}
\caption{Abundances for HD 115444(open squares), HD 186478(crosses), 
HD 108577(triangles) and
BD +8 2548(filled circles).  They have been scaled to match the Arlandini \etal{} (1999)
 solar system
r-process curve using the mean difference between the Sm, Eu, La, Ba
and Ce abundances.  We also show the K\"appeler \etal{} (1989) r-process
curve to illustrate the errors present in deriving solar-system r-process
contributions.  The differences between the two curves arise because
of improved nuclear data and different physical conditions used during
the s-process.  These s-process predictions are then subtracted from 
the total solar-system abundances to produce the curves shown here.}
\end{figure*}

The Th abundance reacts to changes in the model atmosphere
parameters in a very similar manner to the stable neutron-capture elements. 
However, because the Th line is blended, it is possible that the relative 
importance of the contaminants could change because the [Th/Fe] and
[Th/C] ratios do depend on the model atmospheres.   
\twch/\thch{} was determined using  weak \thch{} lines and as a result
the error in the \thch{} contribution is dominated by the error in
the \twch/\thch{} ratio,
rather than the choice of model atmosphere parameters.
As mentioned earlier, the accurate knowledge of [C/H] is not important.
The Fe line does not overlap with the Th line as much as the \thch{} line
and can be monitored independently by looking at the left
wing.  As a result,
error in the [Th/neutron-capture] ratio from the model atomsphere parameters is
surprising small $\sim$ 0.03 dex.  Our total error budget for the
Th abundance includes 0.05 dex for
continuum placement and errors in the contributions of the contaminants
and 0.05 dex due to uncertainties in the \twch/\thch{} ratio.

We have one final note on errors.  Our NdII values appear to be biased
$\sim$ 0.2 dex high in Nd-poor stars. 
 In Nd-rich stars, the two
strongest lines, at 4061.09 \AA{} and 4109.46 \AA, the only ones that
we can measure in the Nd-weak stars, systematically give higher
abundances than the other lines.  We believe that the large
abundances given by the two strongest lines can be traced to errors in
gf values.  The Nd lines in general show large scatter, larger than
can be explained by errors in EW or model atmosphere parameters.
Also, Thevenin (1989), in his compliation of solar gf values, found log gf
values for these two lines that were large by 0.2 dex.  We have chosen
not to correct the Nd-poor stars' values, but caution the reader on the 
accuracy of NdII measurements based on fewer than 3 lines.
  This situation will hopefully
soon be eliminated with the measurement of new NdII gf values.  This strong
line gf problem 
does not appear to affect any of the other elements.

\section{Results}

\subsection{The heavy-element abundance pattern}

The exciting possibility of using Th abundances to estimate ages
of individual stars depends on the reliability of deriving the
initial Th abundance for a star based on the abundance of 
stable elements. The investigations to date suggest a single
(or at least a dominant) ``universal'' r-process abundance pattern,
presumably reflecting the physical conditions in the r-process site
(see earlier discussion).  Although we can only determine a Th
abundance in some of the stars, we can use all the field stars to further
investigate the universality of r-process abundance ratios.  (We did
not use M92 VII-18 for this part of the discussion because of the lower S/N
of its HIRES spectrum and the lack of Hamilton data).  For a qualitative
idea of how well our abundances matched \rss{}, we plotted
the abundances from the four field stars with thorium measurements
and two predictions for the r-process contributions to the
solar system abundances (Figure 3).  
There is good agreement between our
data and \rss, and no obvious s-process contribution, even for the
more metal-rich ([Fe/H]$\sim -2.1$)  

\subsubsection{The s-process contribution}
The onset of substantial s-process contributions to the heavy
elements is the subject of much debate.  CS22892-052 shows no sign of 
s-process contributions for Z$\geq$56
(Sneden 1996).  Neither do HD 115444 ([Fe/H]$\sim -3.0$) or  
HD 122563 ([Fe/H]$\sim -2.7$) (Westin \etal{} 2000).
  McWilliam (1998) measured [Ba/Eu] for 14 stars
with [Fe/H]$ < -2.0$; With the exception of two CH-stars and one star
with [Fe/H]$=-2.07$, this sample showed r-process [Ba/Eu] ratios. He proposes
that only stars more metal-rich than [Fe/H]$\sim -2.0$ show
s-process contributions.  Cowan \etal{} (1996) found that an addition
of 20\% of the total solar system s-process abundances to \rss{} gave
a better fit to the abundances of HD126238 ([Fe/H]$\sim-1.7$)   
Recent models of Galactic chemical evolution (Raiteri \etal{} 1999;
Travaglio \etal{} 1999) predict that s-process contributions to the
Galactic Ba abundance appear at [Fe/H]$\sim-1.7$.

Other studies have found substantial contributions
from the s-process in stars with [Fe/H]$<-2.0$.  Magain (1995)
argued that the profile and width of the BaII line at 4554 \AA{} 
in the subgiant HD 140283 ([Fe/H]$\sim -2.6$) agreed with a dominant contribution from Ba isotopes
produced only in the s-process.  These even isotopes are not affected
by hyperfine splitting, leading to a more pronounced line core.
Mashonkina, Gehren \& Bikmaev (1999) studied the barium abundances 
in cool metal-poor dwarfs.  They found that they could not match
the equivalent widths of the 4554 \AA{} line with the barium abundances
derived from weaker lines unless they adopted a solar ratio of
s- to r-process Ba contributions.  Otherwise the strongest Ba line demands a Ba abundance
that is 0.2-0.3 dex lower than the other lines.  Unfortunately, the same
effects could be mimicked by changes in microturbulent velocity or
changes in temperature structure in the atmosphere, including the
addition of the chromosphere.  We find that using an r-process isotope
distribution decreases the spread in Ba abundances derived from
different lines in our sample of 
giants.  A complicating factor in using the Ba isotopes as s-process vs.
r-process discriminators is the unknown contributions of the even Ba isotopes to
the r-process abundances.  $^{134}$Ba and $^{136}$Ba are blocked from 
contributions to the r-process by stable nuclei; $^{138}$Ba is not.
K\"appeler et al. (1989) found that all $^{138}$Ba can be made in the s-process,
leaving no need for an r-process contribution, 
while Arlandini et al. (1999) found that over half of the Ba made in the r-process
comes from $^{138}$Ba.  Mashonkina et al. (1999) excluded $^{138}$Ba from
their r-process isotope mix.
As discussed in more detail below, if
we look at weaker lines of other elements which are more immune to 
changes in $\xi$ and the temperature structure in the upper layers of the
atmosphere, we confirm that the maximum s-process contribution to even the most
metal-rich star in our sample is $\sim$ 10\%.

Burris et al. (2000) argued on
the basis of [Ba/Eu] ratios for 43 giants that contributions from the s-process
began at [Fe/H]=$-2.9$  
Comparing log $\epsilon$ for the 17 stars we have in common, 
we find a 0.1 dex larger average offset between their
log $\epsilon$(Ba) and ours than we do for the log $\epsilon$(Eu).  While
an offset is expected because of the different model atmospheres, a 
different offset between two rare earth elements is a cause for concern.
As indicated by the previous paragraph, analysis of
Ba is difficult problem, especially the analysis of the strongest
line at 4554\AA{}, which is the only line available for about a third
of the stars in the Burris et al. sample.  Their Ba abundances
were on average 0.21 dex larger than McWilliam (1998) for stars they had
in common.  
\begin{figure*}[htb]
\centerline{
\psfig{figure=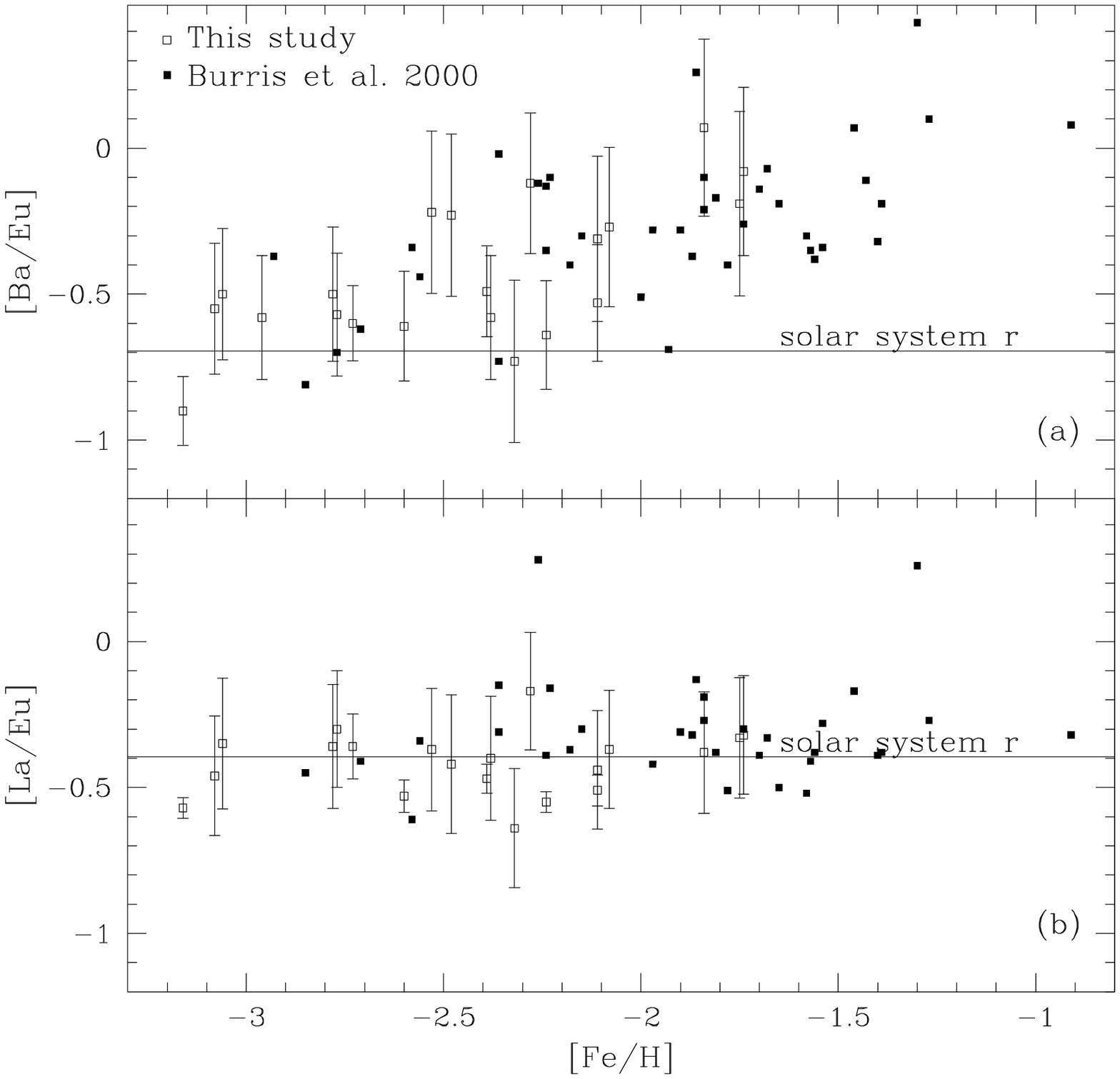,width=7.0truein,angle=0}
}
\caption{(a) [Ba/Eu] vs. [Fe/H] and (b) 
[La/Eu] vs. [Fe/H] for our sample and the Burris et al. (2000) sample.  There
is no trend for increasing [La/Eu] as [Fe/H] increases, as would be predicted
by the [Ba/Eu] results.  We believe that both sets of Ba abundances
become increasingly unreliable as [Fe/H] increases, because of the
strong dependence of Ba on $\xi$.  Our [La/Eu] values are consistent
with only a solar system r-process contribution.  The offset between
the two data sets in [La/Eu] 
is most likely a result of somewhat different linelists
combined with imperfect gf values.  Since we have 17 stars in common,
the offset is not due to markedly different samples of stars.  The
solar system r-process ratios are from Arlandini et al. (1999).  The
poor agreement between the solar system r-process and the lower limit
for our [Ba/Eu] could be due in part to uncertainties in the solar
system predictions (see Figure  3).}
\end{figure*}
\begin{figure*}[htb]
\centerline{
\psfig{figure=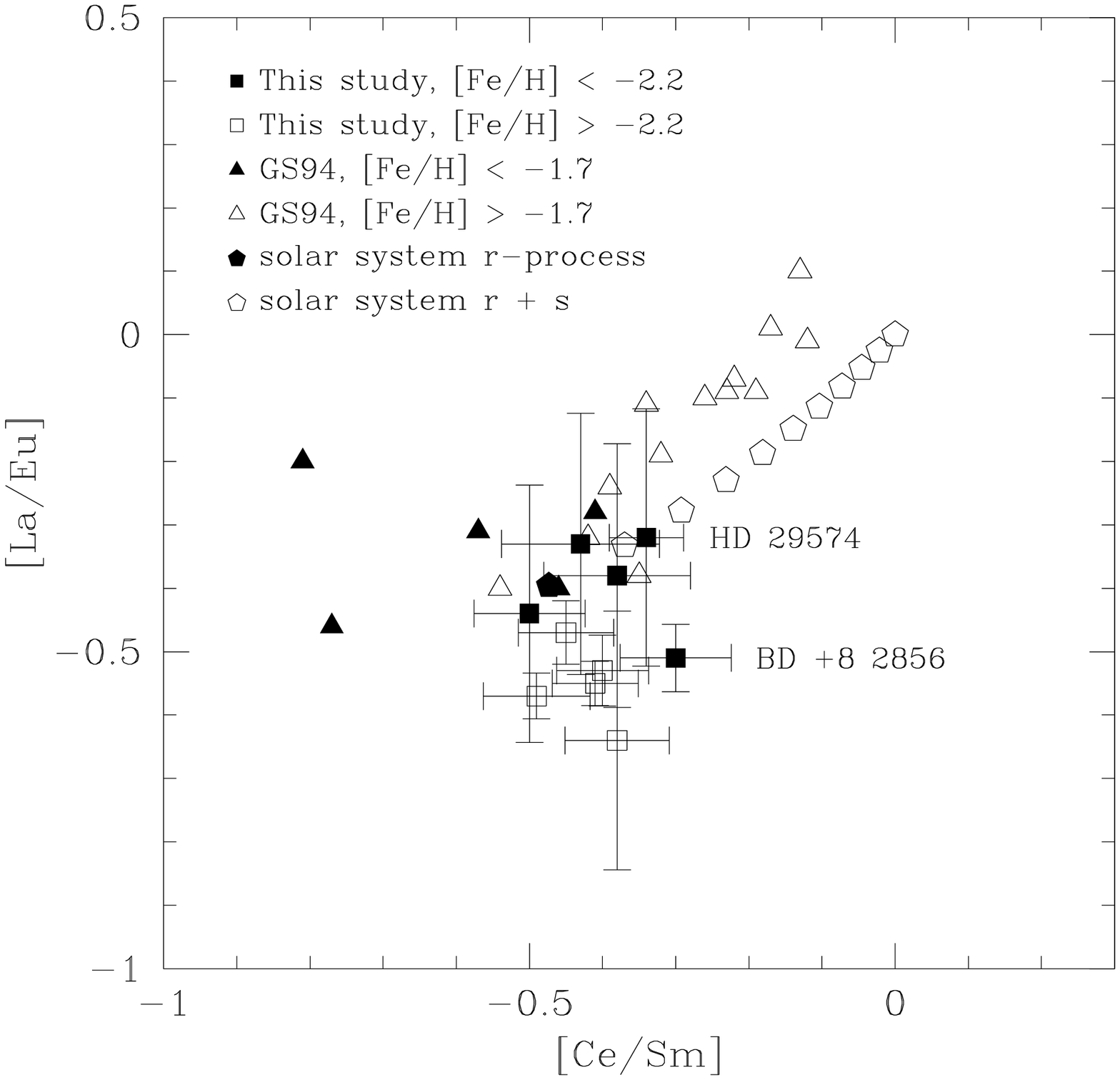,width=7.0truein,angle=0}
}
\caption{Two s-process sensitive ratios, [La/Eu] and
[Ce/Sm].  The filled
pentagon marks the prediction of Arlandini et al. (1999) for
pure r-process abundances.  The open pentagons mark the addition of
s-process material in increments of 10\% of the total solar s-process
abundance.
For our
data, the more metal-rich stars have a higher [La/Eu] due to 
biased Eu values as discussed in the text, but their [Ce/Sm] values
are in agreement with no s-process contribution, with the 
exception of HD29574.  While the Gratton \& Sneden (1994) stars with
[Fe/H]$> -1.7$ follow the path of increasing s-process contributions in spirit
if not in exact numbers, the stars with [Fe/H]$<-1.7$ show no such
inclination.}
\end{figure*}
They attributed this to their smaller microturbulent velocities.
We have plotted in Figure 4 the [Ba/Eu] and [La/Eu]
ratios from both Burris et al. and this study.  
75\% of the solar abundance of La is provided by the s-process,
so this ratio is as sensitive to s-process contributions as
[Ba/Eu].  Figure 4 shows that for both our sample and the Burris
et al. sample, there is no trend toward increasing [La/Eu] at higher
metallicities.  To be consistent with their [Ba/Eu] results, 
[La/Eu] would need to be $\sim$0 for the most metal-rich part of their sample.
We feel the La abundances in both samples are the more robust, and
therefore believe that the [La/Eu] ratios show the true s-process situation.
Finally, although the scatter in our [La/Eu] presented in Figure
4 is consistent with being due to observational errors only,
there could be stars with some small fraction of s-process among our
sample.

We have carried out a simple test with our data using
two s-process sensitive ratios to check for s-process contributions. Figure 5
shows [Ce/Sm] vs. [La/Eu] for the nine stars in our sample that 
have measurements of these
four  elements.  These fall into two groups:  the relatively neutron-capture
rich stars that were used for Th measurements and the more metal-rich
([Fe/H]$ > -2.2$) stars that have measurable Ce and Sm lines.  
We see that for one star (HD 29574, [Fe/H]=$-1.7$) 
an s-process contribution of 10\% could be allowed, while the rest of those with higher
[La/Eu] ratios have [Ce/Sm] ratios
 that agree with the more metal-poor stars.  The
offset between the neutron-capture rich and the metal-rich groups in [La/Eu]
is because the metal-richer stars have their Eu abundances based solely
on the 4129\AA{} line.  The metal-richer stars were preferentially 
observed only with the Hamilton, and the 4129\AA{} was all that was available at reasonable
signal-to-noise in the Hamilton data.  Figure 7 shows that this biases the 
Eu abundances low.  The exception is BD+8 2548, which
is both metal-rich and neutron-capture-rich, 
and has the lowest [La/Eu] of the metal-rich group.
We have also plotted in Figure 5 the data from Gratton \& Sneden (1994) 
who suggested a contribution from the s-process beginning at
[Fe/H]$<-2.0$ as a possible explanation for their heavy element abundances.  
We see
that below [Fe/H] $< -1.7$ their data do not show the trend expected with
an s-process contribution, while the more metal-rich stars 
($-1.7 \leq \rm{[Fe/H]} \leq -0.15$) show the
rising [La/Eu] and [Ce/Sm] ratios that is the signature of increasing
s-process contributions.  There is large scatter and offsets from our
data present in their abundances, especially in the 
metal-poor stars, which we attribute to different, and more uncertain,
gf values used in the Gratton \& Sneden (1994) study.
In summary, we find no convincing evidence for large s-process contributions,
either in our sample, the Burris et al. sample or in the
metal-poor Gratton \& Sneden sample.  For the rest of our analysis, we 
will assume that the heavy-element abundances in our sample of stars represent
contributions from the r-process only.
\begin{figure*}[bht]
\centerline{
\psfig{figure=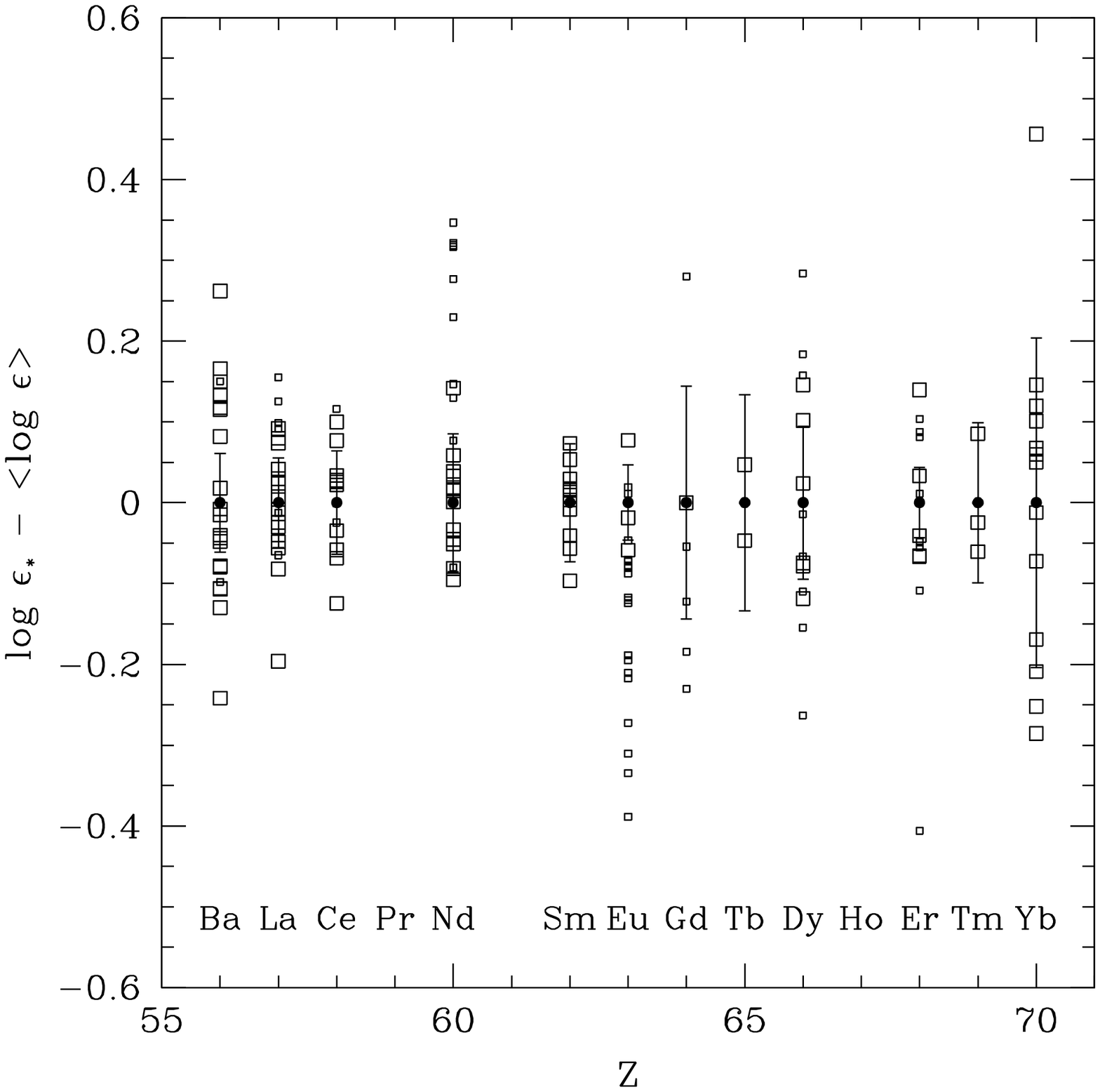,width=5.0truein,angle=0}
}
\caption{$\Delta(\rm log \epsilon)$ between each star's abundance and the mean
abundance for all stars.  Here we have used for the mean calculation 
only measurements based on
3 or more lines, except for Tm, Tb and Yb, which are usually represented,
even in the most heavy-element-rich star, by one line.  These
measurements are plotted with the big squares.  The error bars
represent an average 1-$\sigma$ error bar of the restricted sample 
for an individual data point.  The
error bars are large enough to explain the scatter.  We also plotted the
difference between the mean value and measurements based on fewer than
three lines as small squares.  The abundances of Nd and Eu, in particular,
show the advantage of having multiple measurements.}
\end{figure*}
\subsubsection{The r-process pattern}

Next, we wanted to determine quantitatively if   
all of the metal-poor stars we observed showed a universal r-process pattern in
their heavy element abundances.  
For each
of the stars observed, we used \rss{} as a template to put 
all of the heavy-element abundances in the stars on a common scale.  
 We used  the mean difference
between the Sm, Eu, La, Ba and Ce abundances in the star and the solar 
abundances
attributed by Arlandini et al. (1999) to the r-process to scale each star up, regardless of its [Fe/H] and [heavy-element/Fe] 
values.  We then restricted the sample to abundances that were determined
using three or more lines, with the exception of Tb, Tm and Yb abundances, which,
even in the most favorable cases, were always determined with 1 or 2 lines.
We found the mean value for each element among this scaled, restricted
sample.  The deviation of each star from these mean values
is recorded in Figure 6.  The large symbols indicate measurements which
contributed to the mean, while the small symbols show all the other
deviations.  The larger scatter and bias in the Nd and Eu measurements among
the small symbols show the inherent problems of relying on one or two lines.
\begin{figure*}[htb]
\centerline{
\psfig{figure=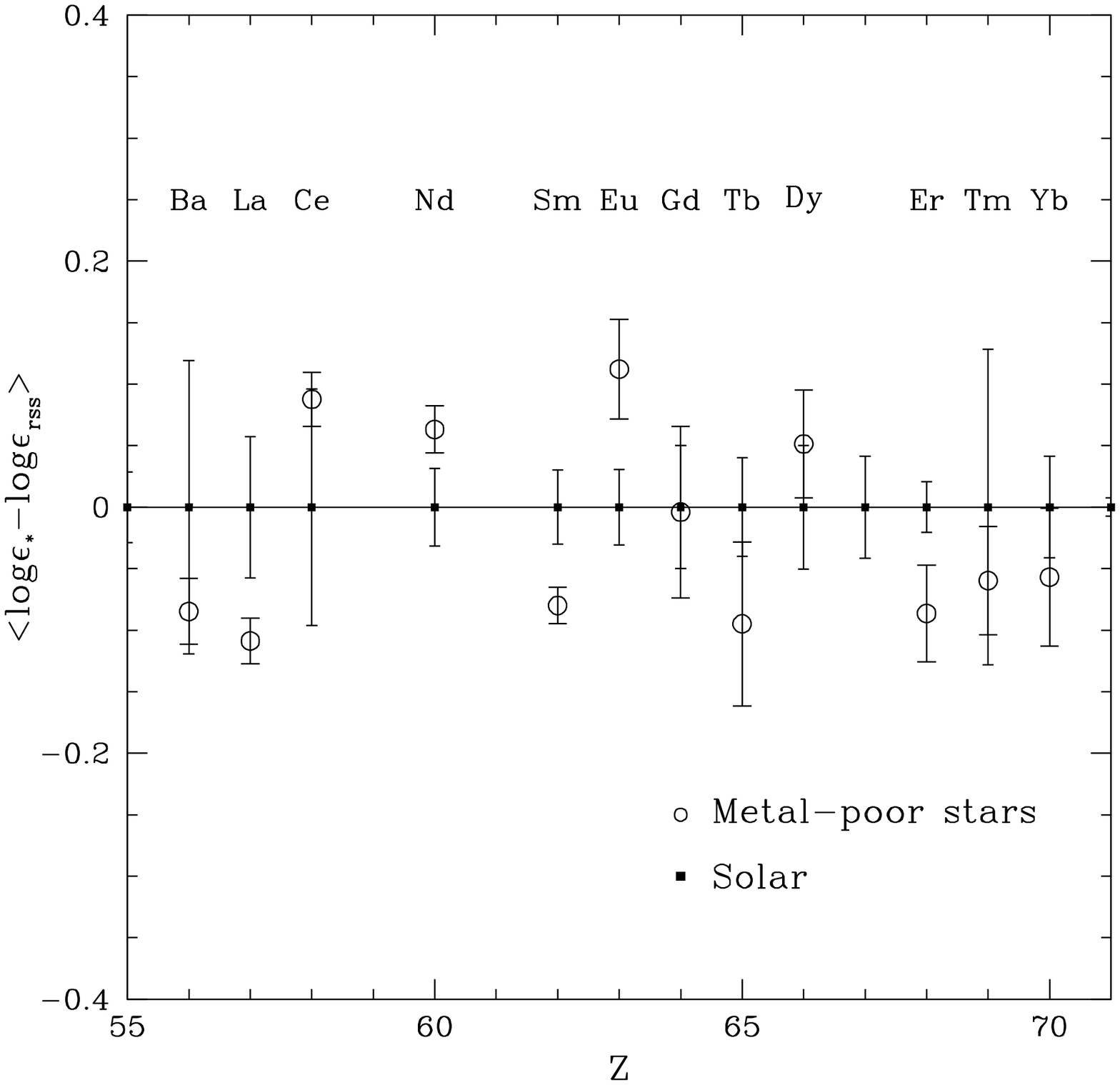,width=5.0truein,angle=0}
}
\caption{A comparison between the abundances in the metal-poor stars and
the r-process contributions to the solar system abundances.  The open
circles (with 1-$\sigma$ error bars) represent the average deviation from the
well-measured abundances in the  
metal-poor
stars from the scaled \rss{}.  There are no differences observed at the
2-$\sigma$ level.}
\end{figure*}  
The first question to ask about the distribution in Figure 6 is whether
the observational errors in the individual points are large enough to account for the
 star-to-star scatter for each of the elements.  We consider
only the random errors and the errors from the overall scaling.  We calculated the latter in a simplistic manner by
finding the average standard deviation of the mean of our five
determinations of the shift for each star.  This was added in
quadrature to the average random error of the restricted sample 
in the abundance determination
to produce the error bars in Figure 6.  In general the error bars are
large enough to account for the observed dispersion.  The exception
might be Ba, but that is the one element for which our assumption that
we could exclude model atmosphere errors is the most unreliable, since
Ba has a large dependence on $\xi$.  Therefore, we argue that
the scatter can be attributed entirely to the observational
errors.

We also calculated the difference between each star's abundance
and \rss{}.  In order to have a true differential comparison
between the abundances in our sample and in the Sun, we used the
total solar abundance derived from the
lines that we could measure in the metal-poor stars, rather than the values
of Anders \& Grevesse (1989).  The r-process fractions were adopted from
Arlandini \etal {} (1999).
\begin{figure*}[htb]
\centerline{
\psfig{figure=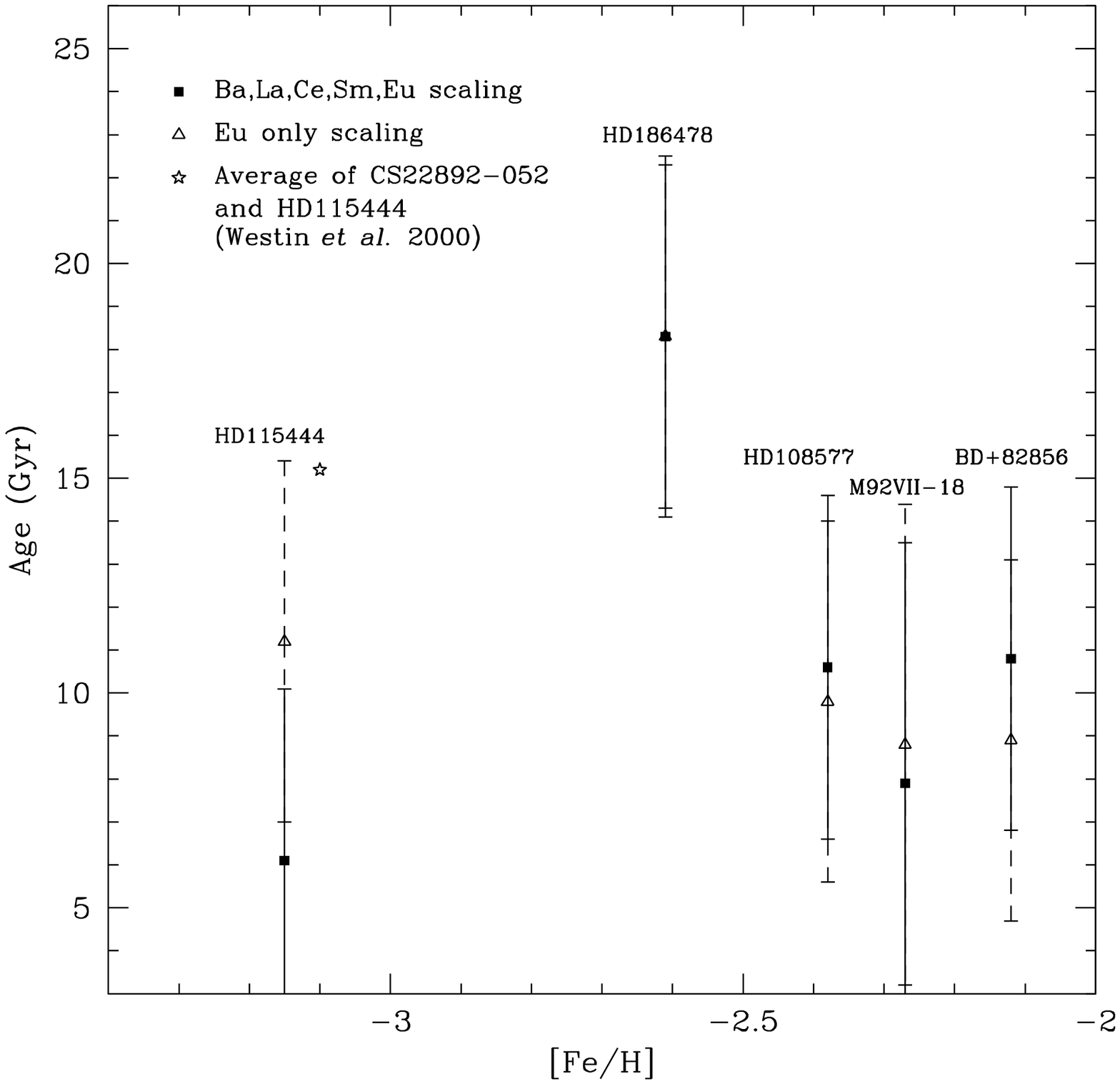,width=5.0truein,angle=0}
}
\caption{Ages from Th abundances.  We have plotted the Case 3 and Case 4
results.  In most cases the two different scalings agree very well; the
exception is HD 115444 which reflects the reality of our 0.05 dex observational
errors in the rare earths.  Also plotted is the average age derived
by Westin \etal{} (2000) from HD 115444 and CS 22892-052.}
\end{figure*}
We plot the mean difference between the restricted sample and \rss, as well
as the s.d.m, in Figure 7 
to determine whether the pattern observed in metal-poor stars is the
same as \rss.  For the errors in the solar values, we include the errors in the solar-system
r-process fraction and in the total solar abundances.  
The fraction of the abundance of an element in
the solar system contributed by the r-process is determined by
subtracting the amount attributed to the s-process from the total
abundance.  While the conditions where the s-process occurs and the
relevant nuclear cross-sections and decay-rates are better known than
for the r-process, errors of 5\% in the predicted s-process abundances
are not uncommon (e.g. K\"appeler, Beer, \& Wisshak 1989).  
For the elements that
are contributed to the solar nebula primarily by the s-process, such as
Ba and La, this can lead to substantial errors in \rss.  In Figure 7,
we use the errors in the solar-system r-process
determinations from Arlandini \etal (1999), which are based on
considering uncertainties in cross-sections and in the physical
conditions where the s-process occurs.  To estimate the uncertainty
in the total solar system abundance, we use the difference between
the photospheric and meteoritic abundances from Anders \& Grevesse (1989),
except for those differences listed as uncertain, in which case
we use 0.04 dex.  Some part of these difference may be attributed to
gf errors, which we have eliminated by using the differential comparison
above.  However, substantial errors also exist because of blending, the
choice of the solar model
atmosphere, damping constants and continuum placement, and we hope to 
provide some idea of those by using the difference between the photospheric
and the meteoritic abundances.
Figure 7 shows no deviations from \rss{} at the 2-$\sigma$ level.  
In summary, our results are
consistent with a single r-process pattern being present in metal-poor
stars and in the Sun from elements with Z$\geq$56.

\subsection{Ages}  
\begin{deluxetable}{cccccccccc}
\tablenum{5}
\tablewidth{0pt}
\tablecaption{Ages from Th Abundances}
\tablehead {
\colhead{Star} & \colhead{log$\epsilon$(Th)} & 
\colhead{log$\epsilon$(Th$_0$)} & \colhead{log$\epsilon$(Th$_0$)}
 & \colhead{log$\epsilon$(Th$_0$)} & \colhead{log$\epsilon$(Th$_0$)} & 
\colhead{Age(Gyr)} & \colhead{Age(Gyr)} & \colhead{Age(Gyr)} & 
\colhead{Age(Gyr)} \\
\colhead{} & \colhead{} & \colhead{(1)} & \colhead{(2)} & \colhead{(3)}
 & \colhead{(4)} & \colhead{(1)} & \colhead{(2)} & \colhead{(3)} & 
\colhead{(4)} }
\startdata
HD 186478 & $-2.26$ & $-1.90$ & $-1.90$ & $-1.87$ & -1.87 & $>16.8$ & $>16.8$ & 
18.3 & 18.3 \nl
HD 115444 & $-2.36$ & $-2.27$ & $-2.15$ & $-2.23$ & $-2.12$ &$>4.2$ & $>9.8$ &
6.1 & 11.2 \nl
HD 108577 & $-1.99$ & $-1.79$ & $-1.81$ &$ -1.76$ & $-1.78$ & $> 9.3$ & $>8.4$ &
 10.6  & 9.8 \nl
BD +8 2548 & $-1.66$ & $-1.46$ & $-1.50$ & $-1.43$ & $-1.47$ & $>9.3$ & $>7.5$ &
 $10.8$ & $8.9$ \nl
M92 VII-18 & $-1.95$ & $-1.81$ & $-1.79$ & $-1.78$ & $-1.76$ & $>6.5$ & $>7.5$
&  $7.9$ & $8.8$ \nl
\enddata
\end{deluxetable}
To find the age of a star from its present Th abundance, we need
to estimate the initial Th abundance.  (We are assuming that the metal
enrichment for these metal-poor stars happened over a short period of
time, so we do not need to model Galactic chemical history).  Given the
results in the previous section justifying the use of the scaling
factor for stable r-process elements for predicting initial Th
abundances we still require an estimate of the original
(Th/stable-r-process) ratio.  There are two possibilities for
estimating this ratio.  The empirical approach is to take the present
abundance of Th in the Sun, correct it for 5 Gyrs of known decay and
use that abundance (log$\epsilon(\rm{Th})_{0,\odot}=0.17$) as a lower limit to
the (Th/stable) production in the r-process.  In reality, of course,
the Sun is made up of gas that has been polluted many times with Th and
stable r-process elements over its history.  The Th has been decaying
over that period, resulting in a smaller (Th/stable) than if it had all
been created 5 Gyrs ago.  In order to turn a lower limit for the age
into an actual value, the Th-production history in the solar
neighborhood must be taken into account (e.g. Cowan et al. 1999).
Also, different stable elements can give different predictions for the
initial Th abundance.  An alternative is to use theoretical
predictions of (Th/stable).  The main disadvantage there is that the
r-process nuclei are descendants of isotopes near the neutron-drip
line, which have very few laboratory measurements of their properties.
Goriely \& Clerbaux (1999) found that the predicted Th/Eu$_0$ ratios
vary from 0.25 to 1.55, depending on which theoretical model for
nuclear properties was used and which solar r-process abundances near
A$=$200 were used as constraints.  If the abundance of $^{209}$Bi was used, the
predicted initial Th/Eu ratio was up to 10 times lower than if the
abundance of $^{206}$Pb was used.  Cowan
\etal{} (1999) regarded the solar r-process abundances as uncertain, and
chose instead to focus on those models which gave reasonable agreement 
with the solar-system r-process abundances over a large range of
A, including A$\sim$200.  
They found that the best fits were given
by three models that gave values of Th/Eu$_{0}$ of 0.496, 0.546
and 0.48.  The situation will be substantially improved as better predictions
for the s-process contribution to the Pb and Bi abundances become
available (e.g. Travaglio \etal{} 2000).  We calculate the ages of five of
the stars in our sample
using both the empirical and the theoretical methods, as well as deriving our initial Th abundances based
on only the Eu abundances versus all of our well-measured heavy
abundances (Table 5).  Eu is produced almost exclusively in the r-process
(only 5\% is due to the s-process), so our derived ages are immune to any
possible s-process contributions.

Case 1
refers to the initial Th estimate obtained by scaling the solar Th
abundance at the time of the solar system formation by the average 
offset between \rss{} and Ba, Ce, Nd, Sm and Eu in the metal-poor stars.  
Case 2 scales the solar 
abundance by the offset between \rss{} and Eu.  Case 3 has the same scaling
as Case 1, but the initial Th abundance is taken from theoretical
predictions (Th/Eu$_{0}$=0.496), while Case 4 takes the Case 3
 initial Th and the
Case 2 scaling.  Figure 8 shows the Case 3 and Case 4 scalings.  
The error in the age depends on the error in the
present Th abundance ($N_{Th}$) and $N_{Th,0}$. 
\begin{equation}
\delta t= \tau \sqrt{\delta (ln N_{Th,0})^2 
+ \delta (ln N_{Th})^2}
\end{equation}
where $\tau$ is the mean life of $^{232}$Th 
\begin{equation}
\tau = 20.3 Gyr  
\end{equation}
Errors in $N_{Th,0}$ are caused by observational errors in the measurement
of stable r-process element abundances and theoretical errors in the Th/Eu
ratio predicted by r-process models.  Right now we will consider only
the random observational errors of 0.05 dex in log$\epsilon$(stable-r)
and the 0.03 dex previously discussed for differential changes in Th and
the rare earths when the atmosphere is changed ($\delta$ log $\epsilon$(Th)$_0$
=0.06).  Errors
in log$\epsilon$(Th) were discussed above and are 0.07 dex in log$(\epsilon_{Th})$,
except for
M92 which has larger observational errors of 0.11 dex.  Using those
errors, we get $\delta t = \pm 4.2$ Gyrs for the field giants and
$\delta t = \pm 5.6$ Gyrs for M92.
We note that our age for HD 115444 using the Ba, Ce, Nd, Sm and Eu scaling is substantially
younger than when using only Eu
scaling.  Figure 3 shows that is due to the low value of Ba we measured in this
star, which produces a lower overall scaling than that suggested by Eu alone.
We can find no obvious mistakes in our Ba abundance, nor is the
dispersion in the abundances derived from the Ba lines particularly
large, but then the offset is only about $-$0.1 dex.

The absolute answer also depends on the log gf value of the
Th line at 4019.12 \AA.  Lawler \etal{} (1990) gave its error
as $\pm 0.04$ dex, which leads to an systematic 
uncertainty in the age of $\pm$ 2 Gyr.  Finally, 
the chosen Th/Eu$_{0}$ is another
source of systematic error.  We summarize our errors, both observational
and theoretical, in Table 6 and show their effect on our age determinations.
\begin{deluxetable}{llc}
\tablewidth{0pt} 
\tablenum{6}
\tablecaption{Summary of Errors}
\tablehead{
\colhead{Cause of Errors} & \colhead{Error} & \colhead{Error
in Gyr}
}
\startdata
\multicolumn{3}{c}{Errors for present $\epsilon$(Th)} \\
$^{12}$CH/$^{13}$CH ratio & 0.05 dex & 2.3 \\
Continuum placement and other blends & 0.05 dex & 2.3 \\
\hspace{.75in}{Total for log$\epsilon$(Th)} & 0.07 dex & 3.0 \\
\multicolumn{3}{c}{Errors for $\epsilon$(Th)$_0$} \\
Changes in Model Atmospheres & 0.03 dex & 1.4 \\
Scatter of log $\epsilon_{r, stable}$ & 0.05 dex & 2.3\\
\hspace{.75in}{Total for log$\epsilon$(Th)$_0$} & 0.06 & 2.8 \\
\multicolumn{3}{c}{Systematic errors} \\
Uncertainties in log gf & 0.04 dex & 2.0 \\
Uncertainties in (Th/Eu)$_{0}$ &  & \\
{\hspace {.5in}} Goriely \& Clerbaux (1999) & $\pm {1.05 \atop 0.28}$ &
$\pm {23.0 \atop 14.2 }$ \\
{\hspace {.5in}} Cowan et al. (1999) & $\pm {0.05 \atop 0.02}$ & $\pm {2.1 
\atop 0.5 }$\\
\enddata
\end{deluxetable}

\section{Discussion}

\begin{deluxetable}{lc}
\tablenum{7}
\tablewidth{0pt}
\tablecaption{Derived Th/Eu$_0$ }
\tablehead{
\colhead{Star} & \colhead{Age} \\
\colhead{} & \colhead{12 Gyr}\\ 
\colhead{} & \colhead{(Th/Eu)$_0$}
} 
\startdata
HD 186478 & 0.36\nl 
HD 115444 & 0.52 \nl 
HD 108577 & 0.53 \nl
BD +8 2548 & 0.57 \nl
M92 VII-18 & 0.57 \nl
\enddata
\end{deluxetable}

The average age for the stars from Case 4 in Table 5 is 11.4 Gyr.  We
chose Case 4 because Case 1 and Case 2 provide only upper limits, 
while Case 3 is affected by the low Ba value of HD 115444 (see above).
Using the Eu scaling also provides a direct comparsion with
theoretical predictions of Th/Eu$_{0}$.  If we vary the  
Th/Eu$_{0}$ from 0.48 to 0.546, our average age ranges from 10.9
to 13.5 Gyr.  All of these values are well within the expansion 
ages derived by Perlmuter \etal{} (1999) and Riess \etal{} (1998).  
We note that in deriving the Th abundances with spectral synthesis, if 
the line list for the Th region is incomplete, we will systematically
overestimate the Th abundance and underestimate the age of the stars. 

For HD115444, Westin \etal{} (2000) found an
age of 15.0 Gyr for (Th/Eu)$_{0}$=0.496, compared with 11.4 Gyr
for this paper.  A comparison of the sytheses of the Th region suggests
the main difference is the CH abundance.  Their synthesis shows
more absorption at 4020 \AA{} than the observed spectrum, which translates
into more \thch{} absorption near the Th line and a smaller Th abundance.
However, a more detailed assessment is not possible because they do
not give the CH and Fe abundances that they used to obtained their fit.
We note that the same \thch/\twch{} ratio was used in both analyses.

Our mean age is based on the assumption of a universal r-process pattern.
Previous studies have shown that when the lighter neutron-capture elements
are considered as well, there is not a consistent pattern 
in different stars. For
example, McWilliam (1998) found variations as large as 2 dex in the
[Sr/Ba] ratio in his sample of metal-poor ([Fe/H]$ < -2.5$  giants.
Westin \etal{} (2000) did a differential comparison of HD 115444 and HD
122563, and also concluded the differences were larger than could be
explained by their observational errors.  Both of these studies,
however, showed a single pattern from Ba to the higher Z elements.  Our
analysis gives a similar result:  regardless of the [heavy-element/Fe]
value, the abundance pattern from Z=56 to Z=70 cannot be distinguished
from \rss.  Goriely \& Arnauld (1997) argued that the agreement between
the \rss{} and metal-poor stars is more a reflection of the underlying
nuclear properties of the elements rather than similarity in conditions
at the r-process site.  Therefore, agreement with a scaled \rss{} over
a limited range does not imply at similar scaling at Z=90.  The
observational results that the third r-process peak elements Os, Pt,
and Ir abundances agree with \rss{} (Sneden \etal{} 1998; Westin
\etal{} 2000) is encouraging in this regard, since that extends the
match with \rss{} over a much larger range in Z.

We can also use our results from a different point of view. If we
assume that all the metal-poor stars for which we have
measured Th are co-eval, we can put a limit
on the observed dispersion in the initial Th/Eu ratio.
Table 7 gives this value assuming all the stars are 12 Gyr old.
Without HD 186478, the range is very small and including it, the
RMS variation is only 0.08. Obviously we have few stars, but
unless there is a large age range in the halo and we have been unfortunate
in our selection of stars, our results support the idea of a single initial
value for the Th/Eu ratio. 
A larger sample of stars would also illuminate
the place of HD 186478 as either a representative of a class that
had a lower Th/Eu$_0$ or as an outlier expected in a statistical
sense.  We emphasize that this low dispersion in Th/Eu$_0$ holds for a particular
sample of stars only -- extremely metal-poor objects which are heavy-element
rich (though the upper limits we have from heavy-element-poor stars
also agree with this limit).

We can also take an age for M92 based on the main-sequence turnoff and
use this to predict (Th/Eu)$_0$. Pont et al. (1998) derive an age of 
14 Gyr, which corresponds to (Th/Eu)$_0$ of 0.63; Carretta et al. (2000)
estimate M92's age at 12.5 Gyr corresponding to 0.57 for (Th/Eu)$_0$.
Both values are within the range of r-process model predictions. This
consistency between different methods of deriving the ages of globular 
clusters is heartening.  Similar results were recently obtained by 
Sneden \etal{} (2000) using three stars in the globular cluster M15.
They found an average age of 14.5$\pm$ 2 Gyrs, again close
to ages derived for the MSTO, assuming (Th/Eu)$_{0}$=0.496
as in this paper. 

The Th-dating of metal-poor stars, in addition to providing lower
limits to the age of the Universe, allows us to examine how the field
stars fit into the overall formation of the Galaxy.  As the sample
size improves, the potential exists to examine whether an age
difference exists between the field stars and the globular clusters
and among the halo field stars themselves.  There is also the
tantalizing possibility that improvements in the accuracy of the age
of the Universe through cosmology and in the ages of the oldest field
stars and globular clusters can result in a star formation history of the
early Milky Way that can be compared with the star formation history of high-z objects.

\acknowledgements
We would like to thank Chris Sneden without whose aid this project would not
have been possible.  We also receive help and guidance from Robert Kraft,
Ruth Peterson and Graeme Smith.  The suggestions by the anonymous
referee greatly improved the paper as well.  J.A.J. acknowledges support from
a NSF Graduate Student Fellowship and a UCSC Dissertation Year Fellowship.
M.B. is happy to acknowledge support from NSF grant AST 94-20204.
Some of the data presented herein was obtained at the W.M. Keck
Observatory, which is operated as a scientific partnership among
the California Institute of Technology, the University of California
and the National Aeronautics and Space Administration.  The Observatory
was made possible by the generous financial support of the W.M. Keck
Foundation.

\clearpage

\begin{deluxetable}{lcccccccccccc}
\tablenum{2A}
\tablewidth{0pt}
\tablecaption{Abundances}
\tablehead{\colhead{Element} &\multicolumn{4}{c}{HD29574} & 
\multicolumn{4}{c}{HD63791} & \multicolumn{4}{c}{HD88609}    \\
\colhead{} & \colhead{[M/Fe]\tablenotemark{a}} & \colhead{$\sigma$} 
& \colhead{$\sigma_{tot}$} & \colhead{N$_{lines}$} & \colhead{[M/Fe]} 
& \colhead{$\sigma$} 
& \colhead{$\sigma_{tot}$} & \colhead{N$_{lines}$} & \colhead{[M/Fe]} 
& \colhead{$\sigma$} 
& \colhead{$\sigma_{tot}$} & \colhead{N$_{lines}$} }
\startdata
FeI  &   $-1.88$ & 0.16 & 0.22 & 151 &    $-1.72$ & 0.16 & 0.21 & 171 &    $-2.97$ & 0.18 & 0.16 & 156 \\
FeII &   $-1.84$ & 0.13 & 0.12 &  15 &    $-1.74$ & 0.14 & 0.16 &  24 &    $-2.96$ & 0.10 & 0.07 &  18 \\
BaII &\phs$ 0.27$ & 0.11 & 0.26 &   4 & \phs$ 0.02$ & 0.07 & 0.26 &   4 &    $-1.09$ & 0.05 & 0.10 &   4 \\
LaII &   $-0.18$ & 0.08 & 0.08 &   4 &    $-0.22$ & 0.05 & 0.14 &   4 &    \nodata & \nodata & \nodata & \nodata \\
CeII &\phs$ 0.09$ & 0.02 & 0.11 &   2 &    $-0.08$ & 0.08 & 0.14 &   5 &    \nodata & \nodata & \nodata & \nodata \\
PrII &   \nodata & \nodata & \nodata & \nodata &    \nodata & \nodata & \nodata & \nodata & $<$$ 0.99$ & \nodata & \nodata & \nodata \\
NdII &\phs$ 0.29$ & 0.21 & 0.12 &  12 & \phs$ 0.11$ & 0.29 & 0.17 &   9 &    $-0.47$ & 0.20 & 0.20 &   1 \\
SmII &\phs$ 0.47$ & 0.25 & 0.13 &   7 & \phs$ 0.26$ & 0.08 & 0.13 &   5 &    \nodata & \nodata & \nodata & \nodata \\
EuII &\phs$ 0.20$ & 0.20 & 0.20 &   1 & \phs$ 0.10$ & 0.20 & 0.24 &   1 &    $-0.51$ & 0.20 & 0.20 &   1 \\
GdII &   \nodata & \nodata & \nodata & \nodata &    \nodata & \nodata & \nodata & \nodata &    \nodata & \nodata & \nodata & \nodata \\
TbII &   \nodata & \nodata & \nodata & \nodata & $<$$ 0.82$ & \nodata & \nodata & \nodata & $<$$ 0.91$ & \nodata & \nodata & \nodata \\
DyII &   \nodata & \nodata & \nodata & \nodata &    \nodata & \nodata & \nodata & \nodata &    \nodata & \nodata & \nodata & \nodata \\
HoII &   \nodata & \nodata & \nodata & \nodata & $<$$ 1.00$ & \nodata & \nodata & \nodata & $<$$ 0.29$ & \nodata & \nodata & \nodata \\
ErII &   \nodata & \nodata & \nodata & \nodata &    \nodata & \nodata & \nodata & \nodata &    $-0.90$ & 0.10 & 0.11 &   1 \\
TmII &   \nodata & \nodata & \nodata & \nodata &    \nodata & \nodata & \nodata & \nodata & $<$$ 0.69$ & \nodata & \nodata & \nodata \\
YbII &   \nodata & \nodata & \nodata & \nodata &    \nodata & \nodata & \nodata & \nodata &    $-1.01$ & 0.20 & 0.21 &   1 \\
HfII &   \nodata & \nodata & \nodata & \nodata &    \nodata & \nodata & \nodata & \nodata &    \nodata & \nodata & \nodata & \nodata \\
OsI  &$<$$ 0.60$ & \nodata & \nodata & \nodata & $<$$ 0.90$ & \nodata & \nodata & \nodata & $<$$ 0.89$ & \nodata & \nodata & \nodata \\
ThII &   \nodata & \nodata & \nodata & \nodata &    \nodata & \nodata & \nodata 
& \nodata &    \nodata & \nodata & \nodata & \nodata \\
\enddata
\tablenotetext{a}{All abundances given as [Element/Fe], except for Fe where
[Fe/H] is given.}
\end{deluxetable}
\begin{deluxetable}{lcccccccccccc}
\tablenum{2B}
\tablewidth{0pt}
\tablecaption{Abundances}
\tablehead{\colhead{Element} &\multicolumn{4}{c}{HD 108577} & 
\multicolumn{4}{c}{HD 115444} & \multicolumn{4}{c}{HD 122563}    \\
\colhead{} & \colhead{[M/Fe]} & \colhead{$\sigma$} 
& \colhead{$\sigma_{tot}$} & \colhead{N$_{lines}$} & \colhead{[M/Fe]} 
& \colhead{$\sigma$} 
& \colhead{$\sigma_{tot}$} & \colhead{N$_{lines}$} & \colhead{[M/Fe]} 
& \colhead{$\sigma$} 
& \colhead{$\sigma_{tot}$} & \colhead{N$_{lines}$} }
\startdata
FeI  &   $-2.38$ & 0.12 & 0.13 & 168 &    $-3.15$ & 0.13 & 0.11 & 149 &    $-2
.7
5$ & 0.16 & 0.15 & 161 \\
FeII &   $-2.39$ & 0.10 & 0.11 &  23 &    $-3.16$ & 0.08 & 0.06 &  19 &    $-2
.7
7$ & 0.11 & 0.08 &  21 \\
BaII &   $-0.10$ & 0.10 & 0.21 &   4 &    $-0.07$ & 0.08 & 0.15 &   4 &    $-1.17$ & 0.01 & 0.11 &   4 \\
LaII &   $-0.08$ & 0.09 & 0.13 &   4 & \phs$ 0.26$ & 0.05 & 0.07 &   4 &    $-0.90$ & 0.10 & 0.11 &   1 \\
CeII &   $-0.22$ & 0.07 & 0.14 &   3 & \phs$ 0.08$ & 0.11 & 0.11 &   3 &    \nodata & \nodata & \nodata & \nodata \\
PrII &$<$$ 0.50$ & \nodata & \nodata & \nodata & \phs$ 0.30$ & 0.20 & 0.15 &   2 & $<$$ 0.50$ & \nodata & \nodata & \nodata \\
NdII &\phs$ 0.06$ & 0.16 & 0.14 &   7 & \phs$ 0.30$ & 0.20 & 0.10 &   8 &    $-0.61$ & 0.20 & 0.20 &   1 \\
SmII &\phs$ 0.23$ & 0.14 & 0.14 &   8 & \phs$ 0.57$ & 0.09 & 0.10 &   7 &    \nodata & \nodata & \nodata & \nodata \\
EuII &\phs$ 0.39$ & 0.02 & 0.12 &   2 & \phs$ 0.83$ & 0.03 & 0.05 &   3 &    $-0.60$ & 0.20 & 0.21 &   1 \\
GdII &\phs$ 0.15$ & 0.63 & 0.46 &   2 & \phs$ 0.57$ & 0.20 & 0.20 &   1 &    \nodata & \nodata & \nodata & \nodata \\
TbII &$<$$ 0.62$ & \nodata & \nodata & \nodata & \phs$ 0.42$ & 0.15 & 0.12 &   2 & $<$$ 0.62$ & \nodata & \nodata & \nodata \\
DyII &\phs$ 0.29$ & 0.13 & 0.11 &   6 & \phs$ 0.76$ & 0.13 & 0.07 &   9 &    \nodata & \nodata & \nodata & \nodata \\
HoII &$<$$ 1.00$ & \nodata & \nodata & \nodata &    \nodata & \nodata & \nodata & \nodata & $<$$ 0.60$ & \nodata & \nodata & \nodata \\
ErII &\phs$ 0.36$ & 0.02 & 0.11 &   3 & \phs$ 0.83$ & 0.04 & 0.04 &   3 &    $-0.60$ & 0.10 & 0.10 &   1 \\
TmII &\phs$ 0.27$ & 0.15 & 0.19 &   1 & \phs$ 0.67$ & 0.15 & 0.11 &   2 & $<$$ 0.30$ & \nodata & \nodata & \nodata \\
YbII &\phs$ 0.20$ & 0.20 & 0.28 &   1 & \phs$ 0.80$ & 0.20 & 0.28 &   1 &    $-1.10$ & 0.20 & 0.20 &   1 \\
HfII &   \nodata & \nodata & \nodata & \nodata &    \nodata & \nodata & \nodata & \nodata & $<$$ 1.00$ & \nodata & \nodata & \nodata \\
OsI  &$<$$ 0.90$ & \nodata & \nodata & \nodata & $<$$ 0.90$ & \nodata & \nodata & \nodata & $<$$ 0.90$ & \nodata & \nodata & \nodata \\
ThII &\phs$ 0.27$ & 0.07 & 0.14 &   1 & \phs$ 0.67$ & 0.07 & 0.09 &   1 &    $
<$$-0.30$ 
& \nodata & \nodata & \nodata \\
\enddata
\end{deluxetable}

\begin{deluxetable}{lcccccccccccc}
\tablenum{2C}
\tablewidth{0pt}
\tablecaption{Abundances}
\tablehead{\colhead{Element} &\multicolumn{4}{c}{HD 126587} & 
\multicolumn{4}{c}{HD 128279} & \multicolumn{4}{c}{HD 165195}    \\
\colhead{} & \colhead{[M/Fe]} & \colhead{$\sigma$} 
& \colhead{$\sigma_{tot}$} & \colhead{N$_{lines}$} & \colhead{[M/Fe]} 
& \colhead{$\sigma$} 
& \colhead{$\sigma_{tot}$} & \colhead{N$_{lines}$} & \colhead{[M/Fe]} 
& \colhead{$\sigma$} 
& \colhead{$\sigma_{tot}$} & \colhead{N$_{lines}$} }
\startdata
FeI  &   $-3.08$ & 0.09 & 0.12 & 137 &    $-2.40$ & 0.12 & 0.14 & 147 &    $-2.32$ & 0.18 & 0.19 & 163 \\
FeII &   $-3.08$ & 0.06 & 0.09 &  18 &    $-2.38$ & 0.11 & 0.12 &  20 &    $-2.32$ & 0.14 & 0.10 &  23 \\
BaII &   $-0.13$ & 0.12 & 0.16 &   4 &    $-0.48$ & 0.03 & 0.17 &   3 &    $-0.24$ & 0.04 & 0.20 &   4 \\
LaII &   $-0.04$ & 0.08 & 0.12 &   3 &    $-0.30$ & 0.15 & 0.16 &   4 &    $-0.15$ & 0.05 & 0.04 &   4 \\
CeII &   \nodata & \nodata & \nodata & \nodata &    \nodata & \nodata & \nodata & \nodata &    $-0.11$ & 0.08 & 0.08 &   3 \\
PrII &$<$$ 1.00$ & \nodata & \nodata & \nodata & $<$$ 1.00$ & \nodata & \nodata & \nodata &    \nodata & \nodata & \nodata & \nodata \\
NdII &\phs$ 0.21$ & 0.20 & 0.18 &   2 & \phs$ 0.10$ & 0.20 & 0.24 &   1 & \phs$ 0.08$ & 0.20 & 0.09 &  12 \\
SmII &   \nodata & \nodata & \nodata & \nodata &    \nodata & \nodata & \nodata & \nodata & \phs$ 0.27$ & 0.14 & 0.08 &   7 \\
EuII &\phs$ 0.42$ & 0.20 & 0.23 &   1 & \phs$ 0.10$ & 0.20 & 0.25 &   1 & \phs$ 0.49$ & 0.20 & 0.20 &   1 \\
GdII &\phs$ 0.24$ & 0.57 & 0.42 &   2 &    \nodata & \nodata & \nodata & \nodata &    \nodata & \nodata & \nodata & \nodata \\
TbII &$<$$ 0.82$ & \nodata & \nodata & \nodata & $<$$ 0.82$ & \nodata & \nodata & \nodata &    \nodata & \nodata & \nodata & \nodata \\
DyII &\phs$ 0.42$ & 0.20 & 0.13 &   3 & \phs$ 0.27$ & 0.20 & 0.20 &   2 & \phs$ 0.66$ & 0.20 & 0.15 &   2 \\
HoII &   \nodata & \nodata & \nodata & \nodata & $<$$ 1.00$ & \nodata & \nodata & \nodata &    \nodata & \nodata & \nodata & \nodata \\
ErII &\phs$ 0.44$ & 0.10 & 0.12 &   2 & \phs$ 0.20$ & 0.10 & 0.17 &   1 &    \nodata & \nodata & \nodata & \nodata \\
TmII &$<$$ 0.90$ & \nodata & \nodata & \nodata & $<$$ 1.00$ & \nodata & \nodata & \nodata &    \nodata & \nodata & \nodata & \nodata \\
YbII &\phs$ 0.15$ & 0.20 & 0.22 &   1 &    $-0.30$ & 0.20 & 0.25 &   1 &    \nodata & \nodata & \nodata & \nodata \\
HfII &   \nodata & \nodata & \nodata & \nodata &    \nodata & \nodata & \nodata & \nodata &    \nodata & \nodata & \nodata & \nodata \\
OsI  &$<$$ 0.90$ & \nodata & \nodata & \nodata & $<$$ 1.00$ & \nodata & \nodata & \nodata & $<$$ 0.60$ & \nodata & \nodata & \nodata \\
ThII &   \nodata & \nodata & \nodata & \nodata &  $<$\phs$0.40$ & \nodata & 
\nodata  & \nodata &    \nodata & \nodata & \nodata & \nodata \\
\enddata
\end{deluxetable}

\begin{deluxetable}{lcccccccccccc}
\tablenum{2D}
\tablewidth{0pt}
\tablecaption{Abundances}
\tablehead{\colhead{Element} &\multicolumn{4}{c}{HD 186478} & 
\multicolumn{4}{c}{HD 216143} & \multicolumn{4}{c}{HD 218857}    \\
\colhead{} & \colhead{[M/Fe]} & \colhead{$\sigma$} 
& \colhead{$\sigma_{tot}$} & \colhead{N$_{lines}$} & \colhead{[M/Fe]} 
& \colhead{$\sigma$} 
& \colhead{$\sigma_{tot}$} & \colhead{N$_{lines}$} & \colhead{[M/Fe]} 
& \colhead{$\sigma$} 
& \colhead{$\sigma_{tot}$} & \colhead{N$_{lines}$} }
\startdata
FeI  &   $-2.61$ & 0.12 & 0.16 & 167 &    $-2.23$ & 0.15 & 0.19 & 165 &    $-2.19$ & 0.11 & 0.16 & 148 \\
FeII &   $-2.60$ & 0.10 & 0.08 &  24 &    $-2.24$ & 0.11 & 0.10 &  25 &    $-2.19$ & 0.13 & 0.14 &  21 \\
BaII &   $-0.08$ & 0.15 & 0.22 &   4 &    $-0.19$ & 0.09 & 0.23 &   4 &    $-0.41$ & 0.20 & 0.24 &   4 \\
LaII &\phs$ 0.01$ & 0.08 & 0.09 &   4 &    $-0.11$ & 0.05 & 0.08 &   4 &    $-0.40$ & 0.10 & 0.17 &   1 \\
CeII &   $-0.09$ & 0.06 & 0.11 &   4 &    $-0.09$ & 0.06 & 0.10 &   4 &    \nodata & \nodata & \nodata & \nodata \\
PrII &$<$$ 0.50$ & \nodata & \nodata & \nodata &    \nodata & \nodata & \nodata & \nodata &    \nodata & \nodata & \nodata & \nodata \\
NdII &\phs$ 0.14$ & 0.19 & 0.11 &  12 & \phs$ 0.14$ & 0.23 & 0.12 &  15 & \phs$ 0.15$ & 0.20 & 0.24 &   1 \\
SmII &\phs$ 0.31$ & 0.17 & 0.12 &   9 & \phs$ 0.32$ & 0.12 & 0.11 &   6 &    \nodata & \nodata & \nodata & \nodata \\
EuII &\phs$ 0.54$ & 0.06 & 0.08 &   3 & \phs$ 0.45$ & 0.01 & 0.05 &   2 &    \nodata & \nodata & \nodata & \nodata \\
GdII &\phs$ 0.43$ & 0.42 & 0.25 &   3 &    \nodata & \nodata & \nodata & \nodata &    \nodata & \nodata & \nodata & \nodata \\
TbII &$<$$ 0.62$ & \nodata & \nodata & \nodata & $<$$ 0.83$ & \nodata & \nodata & \nodata & $<$$ 1.32$ & \nodata & \nodata & \nodata \\
DyII &\phs$ 0.38$ & 0.33 & 0.10 &  13 & \phs$ 0.56$ & 0.20 & 0.22 &   1 &    \nodata & \nodata & \nodata & \nodata \\
HoII &$<$$ 1.00$ & \nodata & \nodata & \nodata &    \nodata & \nodata & \nodata & \nodata &    \nodata & \nodata & \nodata & \nodata \\
ErII &\phs$ 0.53$ & 0.02 & 0.06 &   3 &    \nodata & \nodata & \nodata & \nodata &    \nodata & \nodata & \nodata & \nodata \\
TmII &$<$$ 0.64$ & \nodata & \nodata & \nodata &    \nodata & \nodata & \nodata & \nodata &    \nodata & \nodata & \nodata & \nodata \\
YbII &\phs$ 0.19$ & 0.20 & 0.27 &   1 &    \nodata & \nodata & \nodata & \nodata &    \nodata & \nodata & \nodata & \nodata \\
HfII &$<$$ 1.00$ & \nodata & \nodata & \nodata &    \nodata & \nodata & \nodata & \nodata &    \nodata & \nodata & \nodata & \nodata \\
OsI  &$<$$ 0.60$ & \nodata & \nodata & \nodata &    \nodata & \nodata & \nodata & \nodata & $<$$ 0.90$ & \nodata & \nodata & \nodata \\
ThII &\phs$ 0.23$ & 0.07 & 0.11 &   1 &    \nodata & \nodata & \nodata & 
\nodata
 &    \nodata & \nodata & \nodata & \nodata \\
\enddata
\end{deluxetable}

\begin{deluxetable}{lcccccccccccc}
\tablenum{2E}
\tablewidth{0pt}
\tablecaption{Abundances}
\tablehead{\colhead{Element} &\multicolumn{4}{c}{BD -18 5550} & 
\multicolumn{4}{c}{BD -17 6036} & \multicolumn{4}{c}{BD -11 145}    \\
\colhead{} & \colhead{[M/Fe]} & \colhead{$\sigma$} 
& \colhead{$\sigma_{tot}$} & \colhead{N$_{lines}$} & \colhead{[M/Fe]} 
& \colhead{$\sigma$} 
& \colhead{$\sigma_{tot}$} & \colhead{N$_{lines}$} & \colhead{[M/Fe]} 
& \colhead{$\sigma$} 
& \colhead{$\sigma_{tot}$} & \colhead{N$_{lines}$} }
\startdata
FeI  &   $-3.05$ & 0.10 & 0.08 & 151 &    $-2.77$ & 0.11 & 0.13 & 163 &    $-2.50$ & 0.11 & 0.15 & 135  \\
FeII &   $-3.06$ & 0.12 & 0.08 &  19 &    $-2.78$ & 0.08 & 0.10 &  19 &    $-2.48$ & 0.08 & 0.09 &  21  \\
BaII &   $-0.75$ & 0.15 & 0.16 &   3 &    $-0.45$ & 0.15 & 0.18 &   4 & \phs$ 0.07$ & 0.05 & 0.23 &   4  \\
LaII &   $-0.60$ & 0.10 & 0.14 &   1 &    $-0.30$ & 0.12 & 0.13 &   3 &    $-0.12$ & 0.18 & 0.16 &   2  \\
CeII &   \nodata & \nodata & \nodata & \nodata &    \nodata & \nodata & \nodata & \nodata &    $-0.02$ & 0.11 & 0.13 &   3  \\
PrII &$<$$ 0.65$ & \nodata & \nodata & \nodata & $<$$ 1.01$ & \nodata & \nodata & \nodata &    \nodata & \nodata & \nodata & \nodata  \\
NdII &   \nodata & \nodata & \nodata & \nodata &    $-0.03$ & 0.20 & 0.23 &   1 & \phs$ 0.08$ & 0.20 & 0.17 &   2  \\
SmII &   \nodata & \nodata & \nodata & \nodata &    \nodata & \nodata & \nodata & \nodata &    \nodata & \nodata & \nodata & \nodata  \\
EuII &   $-0.25$ & 0.20 & 0.22 &   1 & \phs$ 0.06$ & 0.20 & 0.23 &   1 & \phs$ 0.30$ & 0.20 & 0.22 &   1  \\
GdII &   \nodata & \nodata & \nodata & \nodata &    \nodata & \nodata & \nodata & \nodata &    \nodata & \nodata & \nodata & \nodata  \\
TbII &$<$$ 0.47$ & \nodata & \nodata & \nodata & $<$$ 0.83$ & \nodata & \nodata & \nodata & $<$$ 1.32$ & \nodata & \nodata & \nodata  \\
DyII &   $-0.33$ & 0.20 & 0.20 &   1 & \phs$ 0.02$ & 0.20 & 0.22 &   1 &    \nodata & \nodata & \nodata & \nodata  \\
HoII &$<$$ 1.65$ & \nodata & \nodata & \nodata & $<$$ 1.51$ & \nodata & \nodata & \nodata &    \nodata & \nodata & \nodata & \nodata  \\
ErII &   $-0.10$ & 0.10 & 0.12 &   1 & \phs$ 0.01$ & 0.10 & 0.14 &   1 &    \nodata & \nodata & \nodata & \nodata  \\
TmII &$<$$ 1.45$ & \nodata & \nodata & \nodata & $<$$ 0.51$ & \nodata & \nodata & \nodata &    \nodata & \nodata & \nodata & \nodata  \\
YbII &   $-0.82$ & 0.20 & 0.21 &   1 &    $-0.24$ & 0.20 & 0.23 &   1 &    \nodata & \nodata & \nodata & \nodata  \\
HfII &$<$$ 1.15$ & \nodata & \nodata & \nodata &    \nodata & \nodata & \nodata & \nodata &    \nodata & \nodata & \nodata & \nodata  \\
OsI  &$<$$ 1.05$ & \nodata & \nodata & \nodata & $<$$ 0.91$ & \nodata & \nodata & \nodata & $<$$ 0.90$ & \nodata & \nodata & \nodata  \\
ThII &  $<$$-0.05$ & \nodata & \nodata & \nodata &    \nodata & \nodata & 
\nodata & \nodata &    \nodata & \nodata & \nodata & \nodata \\
\enddata
\end{deluxetable}

\begin{deluxetable}{lcccccccccccc}
\tablenum{2F}
\tablewidth{0pt}
\tablecaption{Abundances}
\tablehead{\colhead{Element} &\multicolumn{4}{c}{BD +4 2621} & 
\multicolumn{4}{c}{BD +5 3098} & \multicolumn{4}{c}{BD +8 2856}    \\
\colhead{} & \colhead{[M/Fe]\tablenotemark{a}} & \colhead{$\sigma$} 
& \colhead{$\sigma_{tot}$} & \colhead{N$_{lines}$} & \colhead{[M/Fe]} 
& \colhead{$\sigma$} 
& \colhead{$\sigma_{tot}$} & \colhead{N$_{lines}$} & \colhead{[M/Fe]} 
& \colhead{$\sigma$} 
& \colhead{$\sigma_{tot}$} & \colhead{N$_{lines}$} }
\startdata
FeI  &   $-2.52$ & 0.15 & 0.17 &  69 &    $-2.74$ & 0.11 & 0.14 & 159 &    $-2.12$ & 0.17 & 0.19 & 166 \\
FeII &   $-2.53$ & 0.10 & 0.09 &  18 &    $-2.73$ & 0.10 & 0.10 &  20 &    $-2.11$ & 0.15 & 0.11 &  23 \\
BaII &   $-0.82$ & 0.10 & 0.23 &   1 &    $-0.36$ & 0.11 & 0.18 &   4 &    $-0.08$ & 0.08 & 0.24 &   4 \\
LaII &   $-0.96$ & 0.11 & 0.13 &   3 &    $-0.11$ & 0.10 & 0.16 &   1 &    $-0.06$ & 0.04 & 0.09 &   3 \\
CeII &   \nodata & \nodata & \nodata & \nodata &    \nodata & \nodata & \nodata & \nodata &    $-0.08$ & 0.12 & 0.11 &   5 \\
PrII &$<$$ 0.51$ & \nodata & \nodata & \nodata & $<$$ 1.00$ & \nodata & \nodata & \nodata &    $-0.20$ & 0.14 & 0.13 &   2 \\
NdII &   $-0.42$ & 0.20 & 0.18 &   2 & \phs$ 0.11$ & 0.20 & 0.19 &   2 & \phs$ 0.17$ & 0.24 & 0.12 &  10 \\
SmII &   \nodata & \nodata & \nodata & \nodata &    \nodata & \nodata & \nodata & \nodata & \phs$ 0.22$ & 0.19 & 0.11 &  12 \\
EuII &   $-0.59$ & 0.20 & 0.23 &   1 & \phs$ 0.25$ & 0.07 & 0.12 &   2 & \phs$ 0.45$ & 0.04 & 0.06 &   3 \\
GdII &   \nodata & \nodata & \nodata & \nodata &    \nodata & \nodata & \nodata & \nodata & \phs$ 0.15$ & 0.20 & 0.21 &   1 \\
TbII &$<$$ 0.33$ & \nodata & \nodata & \nodata & $<$$ 0.92$ & \nodata & \nodata & \nodata &    $0.27$ & 0.15 & 0.17 &   1 \\
DyII &   \nodata & \nodata & \nodata & \nodata &    $-0.02$ & 0.20 & 0.22 &   1 & \phs$ 0.56$ & 0.26 & 0.12 &   7 \\
HoII &$<$$ 0.01$ & \nodata & \nodata & \nodata &    \nodata & \nodata & \nodata & \nodata & $<$$ 0.90$ & \nodata & \nodata & \nodata \\
ErII &   $-0.49$ & 0.10 & 0.14 &   1 & \phs$ 0.21$ & 0.07 & 0.12 &   3 & \phs$ 0.38$ & 0.01 & 0.09 &   3 \\
TmII &$<$$ 0.51$ & \nodata & \nodata & \nodata & $<$$ 0.80$ & \nodata & \nodata & \nodata & \phs$ 0.28$ & 0.03 & 0.06 &   3 \\
YbII &   $-1.05$ & 0.20 & 0.22 &   1 & \phs$ 0.03$ & 0.20 & 0.25 &   1 & \phs$ 0.22$ & 0.20 & 0.30 &   1 \\
HfII &   \nodata & \nodata & \nodata & \nodata & $<$$ 1.50$ & \nodata & \nodata & \nodata & $<$$ 1.40$ & \nodata & \nodata & \nodata \\
OsI  &$<$$ 0.51$ & \nodata & \nodata & \nodata & $<$$ 1.50$ & \nodata & \nodata & \nodata & $<$$ 0.60$ & \nodata & \nodata & \nodata \\
ThII &  $<$$-0.60$ & \nodata & \nodata & \nodata &    \nodata & \nodata & 
\nodata & \nodata & \phs$ 0.34$ & 0.07 & 0.10 &   1 \\
\enddata
\end{deluxetable}

\begin{deluxetable}{lcccccccccccc}
\tablenum{2G}
\tablewidth{0pt}
\tablecaption{Abundances}
\tablehead{\colhead{Element} &\multicolumn{4}{c}{BD +9 3223} & 
\multicolumn{4}{c}{BD +10 2495} & \multicolumn{4}{c}{BD +17 3248}    \\
\colhead{} & \colhead{[M/Fe]} & \colhead{$\sigma$} 
& \colhead{$\sigma_{tot}$} & \colhead{N$_{lines}$} & \colhead{[M/Fe]} 
& \colhead{$\sigma$} 
& \colhead{$\sigma_{tot}$} & \colhead{N$_{lines}$} & \colhead{[M/Fe]} 
& \colhead{$\sigma$} 
& \colhead{$\sigma_{tot}$} & \colhead{N$_{lines}$} }
\startdata
FeI  &   $-2.29$ & 0.09 & 0.11 & 128 &    $-2.08$ & 0.13 & 0.17 & 157 &    $-2.11$ & 0.12 & 0.13 & 139 \\
FeII &   $-2.28$ & 0.14 & 0.13 &  19 &    $-2.08$ & 0.11 & 0.13 &  23 &    $-2.11$ & 0.09 & 0.14 &  22 \\
BaII &\phs$ 0.03$ & 0.04 & 0.19 &   4 &    $-0.02$ & 0.04 & 0.24 &   4 & \phs$ 0.49$ & 0.09 & 0.26 &   4 \\
LaII &   $-0.02$ & 0.04 & 0.14 &   2 &    $-0.12$ & 0.06 & 0.13 &   4 & \phs$ 0.36$ & 0.06 & 0.14 &   3 \\
CeII &   \nodata & \nodata & \nodata & \nodata &    $-0.17$ & 0.03 & 0.14 &   2 & \phs$ 0.35$ & 0.01 & 0.14 &   2 \\
PrII &   \nodata & \nodata & \nodata & \nodata &    \nodata & \nodata & \nodata & \nodata &    \nodata & \nodata & \nodata & \nodata \\
NdII &\phs$ 0.45$ & 0.20 & 0.24 &   1 & \phs$ 0.23$ & 0.20 & 0.18 &   3 & \phs$ 0.55$ & 0.23 & 0.16 &   8 \\
SmII &   \nodata & \nodata & \nodata & \nodata &    \nodata & \nodata & \nodata & \nodata & \phs$ 0.85$ & 0.15 & 0.16 &   4 \\
EuII &\phs$ 0.15$ & 0.20 & 0.24 &   1 & \phs$ 0.25$ & 0.20 & 0.24 &   1 & \phs$ 0.80$ & 0.20 & 0.24 &   1 \\
GdII &   \nodata & \nodata & \nodata & \nodata &    \nodata & \nodata & \nodata & \nodata &    \nodata & \nodata & \nodata & \nodata \\
TbII &$<$$ 1.32$ & \nodata & \nodata & \nodata & $<$$ 0.82$ & \nodata & \nodata & \nodata & $<$$ 1.32$ & \nodata & \nodata & \nodata \\
DyII &   \nodata & \nodata & \nodata & \nodata &    \nodata & \nodata & \nodata & \nodata & \phs$ 0.88$ & 0.20 & 0.20 &   2 \\
HoII &   \nodata & \nodata & \nodata & \nodata &    \nodata & \nodata & \nodata & \nodata &    \nodata & \nodata & \nodata & \nodata \\
ErII &   \nodata & \nodata & \nodata & \nodata &    \nodata & \nodata & \nodata & \nodata &    \nodata & \nodata & \nodata & \nodata \\
TmII &   \nodata & \nodata & \nodata & \nodata &    \nodata & \nodata & \nodata & \nodata &    \nodata & \nodata & \nodata & \nodata \\
YbII &   \nodata & \nodata & \nodata & \nodata &    \nodata & \nodata & \nodata & \nodata &    \nodata & \nodata & \nodata & \nodata \\
HfII &   \nodata & \nodata & \nodata & \nodata &    \nodata & \nodata & \nodata & \nodata &    \nodata & \nodata & \nodata & \nodata \\
OsI  &$<$$ 1.50$ & \nodata & \nodata & \nodata & $<$$ 0.90$ & \nodata & \nodata & \nodata & $<$$ 1.50$ & \nodata & \nodata & \nodata \\
ThII &   \nodata & \nodata & \nodata & \nodata &    \nodata & \nodata & 
\nodata & \nodata &    \nodata & \nodata & \nodata & \nodata \\
\enddata
\end{deluxetable}

\begin{deluxetable}{lcccccccc}
\tablenum{2H}
\tablewidth{0pt}
\tablecaption{Abundances}
\tablehead{\colhead{Element} &\multicolumn{4}{c}{BD +18 2890} & 
\multicolumn{4}{c}{M92 VII-18} \\
\colhead{} & \colhead{[M/Fe]} & \colhead{$\sigma$} 
& \colhead{$\sigma_{tot}$} & \colhead{N$_{lines}$} & \colhead{[M/Fe]} 
& \colhead{$\sigma$} 
& \colhead{$\sigma_{tot}$} & \colhead{N$_{lines}$} }
\startdata
FeI  &   $-1.73$ & 0.15 & 0.20 & 165 &    $-2.29$ & 0.06 & 0.47 &   9\\
FeII &   $-1.75$ & 0.13 & 0.16 &  21 &    $-2.24$ & 0.15 & 0.36 &   7\\
BaII &\phs$ 0.24$ & 0.17 & 0.29 &   4 &    $-0.39$ & 0.10 & 0.30 &   1\\
LaII &\phs$ 0.10$ & 0.05 & 0.15 &   4 &    $-0.21$ & 0.07 & 0.05 &   4\\
CeII &\phs$ 0.11$ & 0.20 & 0.17 &   5 &    $-0.45$ & 0.09 & 0.11 &   3\\
PrII &   \nodata & \nodata & \nodata & \nodata &    \nodata & \nodata & \nodata & \nodata\\
NdII &\phs$ 0.36$ & 0.20 & 0.16 &  10 &    $-0.03$ & 0.25 & 0.12 &   5\\
SmII &\phs$ 0.54$ & 0.13 & 0.15 &   5 & \phs$ 0.11$ & 0.20 & 0.16 &   3\\
EuII &\phs$ 0.43$ & 0.20 & 0.24 &   1 & \phs$ 0.30$ & 0.09 & 0.07 &   2\\
GdII &   \nodata & \nodata & \nodata & \nodata & \phs$ 0.44$ & 0.20 & 0.22 &   1\\
TbII &$<$$ 0.82$ & \nodata & \nodata & \nodata &    \nodata & \nodata & \nodata & \nodata\\
DyII &\phs$ 0.49$ & 0.20 & 0.18 &   3 & \phs$ 0.04$ & 0.20 & 0.22 &   1\\
HoII &   \nodata & \nodata & \nodata & \nodata &    \nodata & \nodata & \nodata & \nodata\\
ErII &   \nodata & \nodata & \nodata & \nodata & \phs$ 0.33$ & 0.20 & 0.18 &   2\\
TmII &   \nodata & \nodata & \nodata & \nodata &    \nodata & \nodata & \nodata & \nodata\\
YbII &   \nodata & \nodata & \nodata & \nodata &    $-0.06$ & 0.20 & 0.24 &   1\\
HfII &   \nodata & \nodata & \nodata & \nodata &    \nodata & \nodata & \nodata & \nodata\\
OsI  &$<$$ 0.90$ & \nodata & \nodata & \nodata &    \nodata & \nodata & \nodata & \nodata\\
ThII &   \nodata & \nodata & \nodata & \nodata &    \phs$0.20$ & 0.07 & 0.07 &
 1 \\
\enddata
\end{deluxetable}

\end{document}